\documentclass[10pt]{article}
\usepackage{flushend, cuted}
\usepackage{amsmath}
\usepackage{amsfonts}
\usepackage{bbm}
\usepackage{cite}

\usepackage{color,amsmath,amssymb, amsfonts,amstext,amsthm}

\usepackage{epsfig, graphicx, graphics}

\usepackage{amssymb}
\usepackage{float}

\usepackage{color,amsmath,amssymb, amsfonts,amstext,amsthm}

\usepackage{epsfig, graphicx, graphics}
\numberwithin{figure}{section}
\numberwithin{equation}{section}
\numberwithin{table}{section}

\usepackage{longtable}
\usepackage{cite}
\usepackage{subfigure}
\usepackage{diagbox}
\usepackage{authblk}
\usepackage{ragged2e}
\renewcommand{\phi}{\varphi}

\title{Detecting the most probable transition phenomenon of a nutrient–phytoplankton–zooplankton system}

\author[1]{Hui Wang}
\author[1]{Ying Wang\footnote{w281936522@163.com}}
\author[2]{Xi Chen}

\affil[1]{School of Mathematics and Statistics, Zhengzhou University\\100 Kexue Road, Zhengzhou 450001, China}
\affil[2]{School of Statistics, Xi 'an University of Finance and Economics, Xi'an 710100, China}

\date{\today}

\begin{document}

\maketitle

\begin{abstract}
The population biology model holds a significant position within ecosystems. Introducing stochastic perturbations into the model can more accurately depict real biological processes. In this paper, we primarily investigate the most probable transition phenomenon in a three-dimensional nutrient-phytoplankton-zooplankton (NPZ) plankton model. With appropriate parameter values, the system coexists with a stable equilibrium point and a stable limit cycle. Under noise perturbations, transitions occur between these two steady states. Based on the Onsager-Machlup action functional and the neural shooting method, we have studied the most probable transition time, the most probable transition pathway and the most probable transition probability of the NPZ system. The transition between these metastable states plays a crucial role in stochastic ecosystems, providing guidance for a better understanding of complex biological processes.

\end{abstract}

\section{Introduction}
\indent

Stochastic dynamical systems represent a significant branch of the theory of dynamical systems. They integrate deterministic dynamical systems with stochastic noise to describe the random evolution of system states. Stochastic noise is typically represented by Gaussian noise and non-Gaussian noise \cite{Duan}. The incorporation of stochastic noise makes dynamical systems more reflective of phenomena in real world. Stochastic dynamical systems have extensive applications in biology, medicine, engineering and other fields \cite{Bressloff2021Stochastic,Bashkirtseva2016,Neu2016}.

Stochastic dynamical systems are widely applied across various fields. In biomedicine, researchers have investigated the impact of stochastic effects on cardiac rhythm analysis \cite{Cheffer2020heart}. In climate modeling, it has been observed that climate systems can exhibit mixed-mode stochastic oscillations near equilibrium \cite{Alexandrov2016climate}. In economics, a novel securities pricing model featuring jump and feedback-driven stochastic volatility has been proposed in \cite{Sun2006price}. In quantum mechanics, Ji and colleagues have delved into fundamental quantum stochastic processes and related quantum stochastic integrals \cite{Ji2016QSC}. Moreover, stochastic dynamical systems have also been extensively investigated in ecology \cite{Gottesman2012eco,Kamenev2008eco}. This broad application of stochastic dynamical systems underscores their importance in advancing our understanding across diverse scientific domains.

Population biological models are crucial tools for studying the dynamics of biological populations. These models use mathematical frameworks to describe the structural characteristics of populations as well as their interactions with other organisms and the environment. When resources are limited, population growth follows the Logistic model \cite{Han2017Logistic}. Meanwhile, the Lotka-Volterra model is employed to depict the interactions between predators and prey \cite{Roy1987Lotka}. These models provide a theoretical foundation for understanding the relationships among different populations. The ecological relationships among populations within marine systems have also garnered the attention of scientists. The nutrient-phytoplankton-zooplankton (NPZ) model is a vital tool for studying the dynamics of plankton in marine ecosystems, holding significant ecological importance. By simulating the interactions among nutrients, phytoplankton, and zooplankton, it helps elucidate the fundamental principles of material cycling and energy flow within marine ecosystems. Phytoplankton, as primary producers, convert nutrients into organic matter through photosynthesis, thereby providing the foundation for the entire marine food chain. In this process, phytoplankton absorb nutrients via photosynthesis, while zooplankton feed on phytoplankton, forming the initial link in the food chain. These interactions not only influence the population dynamics of plankton, but also have far-reaching effects on the stability and biodiversity of the entire ecosystem. Moreover, the NPZ model plays a crucial role in investigating the dynamic changes of ecosystems. By simulating the growth and reproduction of plankton under various environmental conditions, the model can reveal the mechanisms by which ecosystems respond to environmental changes, such as climate change and eutrophication. In summary, as an important ecological tool, the NPZ model not only aids in our in-depth understanding of the functioning of marine ecosystems but also holds significant application value in predicting ecological disasters, protecting the marine environment and managing fishery resources.

The origins of the NPZ model can be traced back to the mid-20th century, when scientists began to focus on the dynamics of plankton in marine ecosystems and their interactions with nutrients. Early research primarily concentrated on the growth of phytoplankton and the cycling of nutrients. However, as studies progressed, the role of zooplankton was also incorporated into the models. A significant milestone in the development of the NPZ model came in $1962$ when Steele proposed a simple model for phytoplankton growth, focusing on how phytoplankton absorb nutrients through photosynthesis and convert them into biomass \cite{Steele1962}. The model took a major step forward in 1981 when Steele and Henderson introduced a three-component model that used differential equations to describe the interactions among nutrients, phytoplankton and zooplankton \cite{SH1981}. This pioneering model laid the foundation for subsequent research. Building on this foundation, Edwards and Brindley enhanced the NPZ model by incorporating the effects of light and temperature on phytoplankton growth, as well as the nonlinear relationships of zooplankton grazing efficiency and respiration \cite{Edwards1996NPZ}. Huisman and Weissing focused on the vertical migration of phytoplankton and the distribution of light, highlighting their impact on ecosystem dynamics \cite{Huisman1999}. Steele and Henderson demonstrated that the functional form of zooplankton predation could affect the stability of the model in 1993 \cite{SH1993}. In 2005, Lampert further refined the model by including zooplankton behavior, such as vertical migration, thereby providing a more comprehensive description of ecosystem dynamics \cite{Lampert2005}. These advancements have continuously enriched the NPZ model, making it a powerful tool for understanding the complex interactions within marine ecosystems and their responses to various environmental factors.

The Onsager–Machlup action functional theory is a crucial tool for investigating transition phenomena in stochastic dynamical systems. It employs the path integral approach to depict the most probable pathway of a system transitioning from one state to another. Initially proposed by Onsager and Machlup in 1953 \cite{Onsager1953OM}, the Onsager-Machlup action functional was designed to describe diffusion processes with linear drift coefficients and constant diffusion coefficients. The core concept of the Onsager-Machlup action functional is to express the probability density of a path as a functional integral, thereby identifying the most probable pathway within the system \cite{Liu2024OM}. The theory of the Onsager–Machlup action functional has been widely applied across various fields. In stochastic Hamiltonian systems, the Onsager-Machlup action functional is utilized to determine the most probable transition pathways among system trajectories \cite{Zhang2025OMHamilton}. It has also been employed to study fluctuation phenomena in dynamical processes, such as the motion of Brownian particles and the dynamical behavior in complex fluids \cite{Taniguchi2007Fluc}. Furthermore, the Onsager-Machlup action functional theory has been extended to non-equilibrium states, leading to the establishment of several fluctuation theorems \cite{Deza2009noneq}. Researchers also have investigated the applications of the Onsager-Machlup action functional in elliptic diffusion processes and jump-diffusion processes \cite{ellipticOM,Chao2019OMJump}. These studies have significantly contributed to the theoretical understanding of transition phenomena in stochastic dynamical systems. In addition, the neural shooting method has garnered considerable attention from researchers in recent years. Jean-Pierre Sleiman and his colleagues have combined neural networks with the iterative Linear Quadratic Regulator (iLQR) method and have effectively addressed trajectory optimization problems \cite{Cheng2022-ilQR}. For data that exhibit oscillatory behavior or span long time intervals, the multiple shooting method can successfully fit them \cite{Turan2022shoot}. Du et al. \cite{Du2024spectral} have proposed a neural network method based on the spectral domain, and it enables more efficient differentiation within neural networks and substantially reduces the complexity of training. In summary, the neural shooting method has a wide range of applications and occupies a significant position in stochastic dynamical systems.

The NPZ system illustrates the cycling of nutrients between phytoplankton and zooplankton. The Holling Type III functional response models the selective feeding behavior of zooplankton on phytoplankton, such as the escape effect or threshold feeding \cite{Holling1959Type,Oaten1975Type}. Moreover, the Type III functional response can also inhibit the excessive growth of phytoplankton and prevent the system from becoming uncontrollable, such as the occurrence of red tides. Additionally, the growth of phytoplankton is regulated by both nutrient availability and its own population density \cite{Edwards1999}. The external input of nutrients and the mortality of zooplankton play a crucial role in the stability of the system. We primarily focus on the stability of the system and the most probable transition phenomena. The stable equilibrium point and the stable limit cycle correspond to the steady state of low nutrient levels and the steady state of periodic oscillation, respectively. For example, the stable equilibrium point may represent a clear water state, while the stable limit cycle may correspond to the cycle of seasonal outbreaks of phytoplankton and predation by zooplankton. Through stability analysis, we can determine whether phytoplankton and zooplankton will collapse due to over-predation or achieve balance through self-regulation. Moreover, the system may suddenly switch between these two steady states due to external disturbances, such as shifting from a clear water state to an algal bloom state, thereby causing an irreversible ecological phase transition and increasing the uncertainty of the ecosystem \cite{Behrenfeld2014}. This also explains why different water bodies may exhibit completely different community structures under similar environmental conditions. By studying the stability and most probable transition phenomena of the NPZ system, we can not only deepen our understanding of the dynamic patterns of plankton ecosystems but also provide guidance for ecological management and protection, thereby promoting the sustainable management of ecosystems.

In this paper, we delve into a stochastic nutrient-phytoplankton-zooplankton model. By examining the eigenvalues of the Jacobian matrix, we analyze the stability of the system. Notably, under appropriate parameter values, the system exhibits a fascinating phenomenon: a stable limit cycle and a stable equilibrium point coexist. Leveraging numerical simulation of Brownian motion, we have successfully captured the distribution of the most probable transition times. Capitalizing on the neural shooting method, we have delved into the analysis of the most probable transition pathways within the NPZ system. Based on the Onsager-Machlup action functional theory, we can elucidate the intricate relationship between the most probable transition probability and the value of the action functional. The noise-induced transitions between metastable states are of great significance in ecosystems, offering a novel perspective for understanding the dynamic changes in biological population numbers within these systems.

This paper is organized as follows. In Section 2, we introduce a three-dimensional NPZ system with stochastic noise and analyze the stability of the system. In Section 3, we present methods for simulating Brownian motion, the Onsager-Machlup action functional, as well as the neural shooting method. In Section 4, we unveil our core results, which encompass the most probable transition time, the most probable transition pathway and the most probable transition probability. We conclude our paper with a discussion in Section 5.

\section{The stochastic model}
\indent

The NPZ model is a mathematical framework that describes the interactions among nutrients, phytoplankton, and zooplankton within marine ecosystems. The depletion of nutrients is driven by the uptake of nutrients by phytoplankton, while the replenishment of nutrients is due to the respiration and excretion of phytoplankton, as well as the excretion of zooplankton and mixing processes. Phytoplankton grow by absorbing nutrients, but their populations are reduced through respiration, grazing by zooplankton, and sedimentation. Zooplankton, on the other hand, grow by preying on phytoplankton, but they also experience natural mortality and predation by higher trophic levels. We assume that the water layer in which the plankton reside is well-mixed, with no concentration gradients. The model does not account for spatial variations, focusing solely on temporal changes. Additionally, the model simplifies the interactions among planktonic organisms, considering only the fundamental interactions among nutrients, phytoplankton, and zooplankton.

\subsection{The deterministic model}
\indent

We consider a three-dimensional NPZ model. Based on the aforementioned assumptions, the changes observed in the three species are described as follows:
\begin{equation}\label{eq_deter}
\begin{cases}
    \frac{dx}{dt}=-\frac{x}{e+x} \frac{a}{b+cy} y+r y+\frac{\beta \lambda y^2}{\mu^2+y^2} z+\gamma d z^2+k(N_0-x), \\
    \frac{dy}{dt}=\frac{x}{e+x} \frac{a}{b+cy} y-r y-\frac{\lambda y^2}{\mu^2 +y^2} z-(s+k) y,   \\
    \frac{dz}{dt}=\frac{\alpha \lambda y^2}{\mu^2 +y^2} z-d z^2, 
\end{cases}
\end{equation}
where $x$, $y$, $z$ denote the concentration of nutrient, phytoplankton and zooplankton respectively, $a$ gives maximum growth rate, $b$ is light attenuation by water, $c$ is phytoplankton self-shading coefficient, $d$ is higher predation on zooplankton, $e$ is half-saturation constant for nutrient uptake, $k$ represents cross thermocline exchange rate, $r$ and $s$ are phytoplankton respiration and sinking loss rate, $\alpha$ and $\beta$ are zooplankton assimilation efficient and excretion fraction, $\gamma$ represents regeneration of $z$ predation excretion, $\lambda$ is maximum grazing rate of $z$, $\mu$ is $z$ grazing half-saturation coefficient, and $N_0$ represents $N$ concentration below mixed layer.

For general differential equations, we can determine the stability of the system by calculating the eigenvalues at the equilibrium points. If all the eigenvalues have negative real parts, the equilibrium point is asymptotically stable. Conversely, if at least one eigenvalue has a non-negative real part, the equilibrium point is unstable \cite{Perko2001SDE}.

The NPZ system (\ref{eq_deter}) has the following possible steady states:

$(1)$ There is only a steady state $(N_0,0,0)$. In this case, the Jacobian matrix of (\ref{eq_deter}) is 
$$\begin{pmatrix}
 -k & -\frac{a N_0}{(e+N_0) b}+r & 0 \\
 0 & \frac{a N_0}{(e+N_0) b}-r-s-k & 0 \\
 0 & 0 & 0 
\end{pmatrix}.$$
It has two eigenvalues $-k$ and $\frac{a N_0}{(e+N_0) b}-r-s-k$. We can conclude that $(N_0,0,0)$ is stable for $\frac{a N_0}{(e+N_0) b}-r-s-k<0$, and is unstable for $\frac{a N_0}{(e+N_0) b}-r-s-k>0$.

$(2)$ There exists two steady states $(x^*,y^*,0)$. $x^*$ and $y^*$ satisfy the following equations:
\begin{equation}\label{case2}
\begin{cases}
    -\frac{x}{e+x} \frac{a}{b+c y} y+r y+k(N_0-x)=0, \\
   \frac{x}{e+x} \frac{a}{b+c y} y-r y-(s+k)y=0,   \\ 
\end{cases}
\end{equation}
then we can obtain the relationship between $x^*$ and $y^*$:
$$k(N_0-x^*)=(s+k)y^*,$$
Substituting this into (\ref{case2}), and eliminating $x$. Then $y^*$ should satisfy the quadratic equation
$$k(s+k)(e+N_0)(s+k+r)c y^2+(k(e+N_0)(s+k)(s+k+r)b+(s+k)a) y-akN_0=0.$$
The discriminant of the quadratic is:
$$\Delta=(k(e+N_0)(s+k)(s+k+r)b+(s+k)a)^2 + 4 k(s+k)(e+N_0)(s+k+r)c akN_0.$$

It is obvious that it is strictly positive, so there are two solutions $y_1^*$ and $y_2^*$. The constant term of the quadratic $-akN_0$ is negative. Hence one of the solutions is negative while the other is always positive (assume $y_1^*<0$ and $y_2^*>0$). The concentration never become negative. So we consider the form of $(x_2^*,y_2^*,0)$. According to the expression of $x^*$ and $y^*$, it is clear that $x_2^*<0$ for $k N_0-(s+k)y_2^*<0$, and $x_2^*>0$ for $k N_0-(s+k)y_2^*>0$.

The Jacobian matrix of $(x^*,y^*,0)$ is 
$$\begin{pmatrix}
 -\frac{y a e}{(b+cy)(e+x)^2}-k & -\frac{x a b}{(e+x)(b+cy)^2}+r+\frac{2 \beta \lambda z y \mu^2}{(\mu^2+y^2)^2} & \frac{\beta \lambda y^2}{\mu^2+y^2}\\
 \frac{e a y}{(e+x)^2(b+cy)} & \frac{x a b}{(e+x)(b+cy)^2}-r-s-k & -\frac{\lambda y^2}{\mu^2+y^2} \\
 0 & 0 & \frac{\alpha \lambda y^2}{\mu^2+y^2} 
\end{pmatrix}.$$
Due to one of the eigenvalues $\frac{\alpha \lambda y^2}{\mu^2+y^2}$ always being positive, the equilibrium point $(x_2^*,y_2^*,0)$ is unstable.

$(3)$ It is clear that the steady state $(x^*,y^*,z^*)$ also exists, where $z^*$ should satisfy
$$z^*=\frac{\alpha \lambda y^{*2}}{d(\mu^2+y^{*2})}.$$
The Jacobian matrix of $(x^*,y^*,z^*)$ is
$$\begin{pmatrix}
 -\frac{y a e}{(b+cy)(e+x)^2}-k & -\frac{x a b}{(e+x)(b+cy)^2}+r+\frac{2 d^2 z^3 \beta \mu^2}{\alpha^2 y^3 \lambda} & \frac{\beta d z}{\alpha}+2 \gamma d z\\
 \frac{e a y}{(e+x)^2(b+cy)} & \frac{x a b}{(e+x)(b+cy)^2}-\frac{2 d^2 z^3 \mu^2}{\alpha^2 y^3 \lambda}-r-s-k & -\frac{d z}{\alpha} \\
 0 & \frac{2 d^2 z^3 \mu^2}{\alpha \lambda y^3} & -d z 
\end{pmatrix}.$$

When the three populations coexist, the equilibrium point is identified as $E=(0.2258, 0.0469, 0.0690)$. By examining the Jacobian matrix at point $E$ 
$$\begin{pmatrix}
    -0.0697 & -0.4540 & 0.2235 \\
    0.0197 & 0.0922 & -0.3854 \\
    0 & 0.1014 & -0.0963
\end{pmatrix},$$
we have calculated that the eigenvalues are $-0.0697$, $-0.0359$ and $-1.0063$, all of which are negative. This suggests that $E$ is a stable equilibrium point. For the stability of a limit cycle, we can use Lyapunov exponents as a criterion. If all Lyapunov exponents are negative, the limit cycle is stable; if the largest Lyapunov exponent is positive, the limit cycle is unstable. In the case of our system (\ref{eq_deter}), the phase diagram indicates the presence of a limit cycle. The Lyapunov exponents for our system are $-0.00058871$, $-0.00585242$, $-0.00785613$. Consequently, we can conclude that when the parameter $d$ is $1.396$, the limit cycle is stable.

Based on the above stability analysis, we choose the following parameter values as the same in \cite{Edwards1996NPZ}: $a=0.2$, $b=0.2$, $c=0.4$, $d=1.396$, $e=0.03$, $k=0.05$, $r=0.15$, $s=0.04$, $N_0=0.6$, $\alpha=0.25$, $\beta=0.33$, $\gamma=0.5$, $\lambda=0.6$, $\mu=0.035$.

In this paper, we consider the NPZ model as depicted in Eq.(\ref{eq_deter}), which not only captures the interactions among nutrients, phytoplankton and zooplankton, but also incorporates external inputs and losses within the system. Additionally, Chen investigated a similar NPZ model in \cite{Chen2024transition}. Unlike our system (\ref{eq_deter}), however, the model focuses more on the relationship among three populations, offering a more simplified representation. Both NPZ models aim to enhance our understanding and prediction of marine ecosystem behavior, including the dynamics of biological populations and their responses to environmental changes.

\begin{figure}[ht]
    \centering
    \includegraphics[height=12cm]{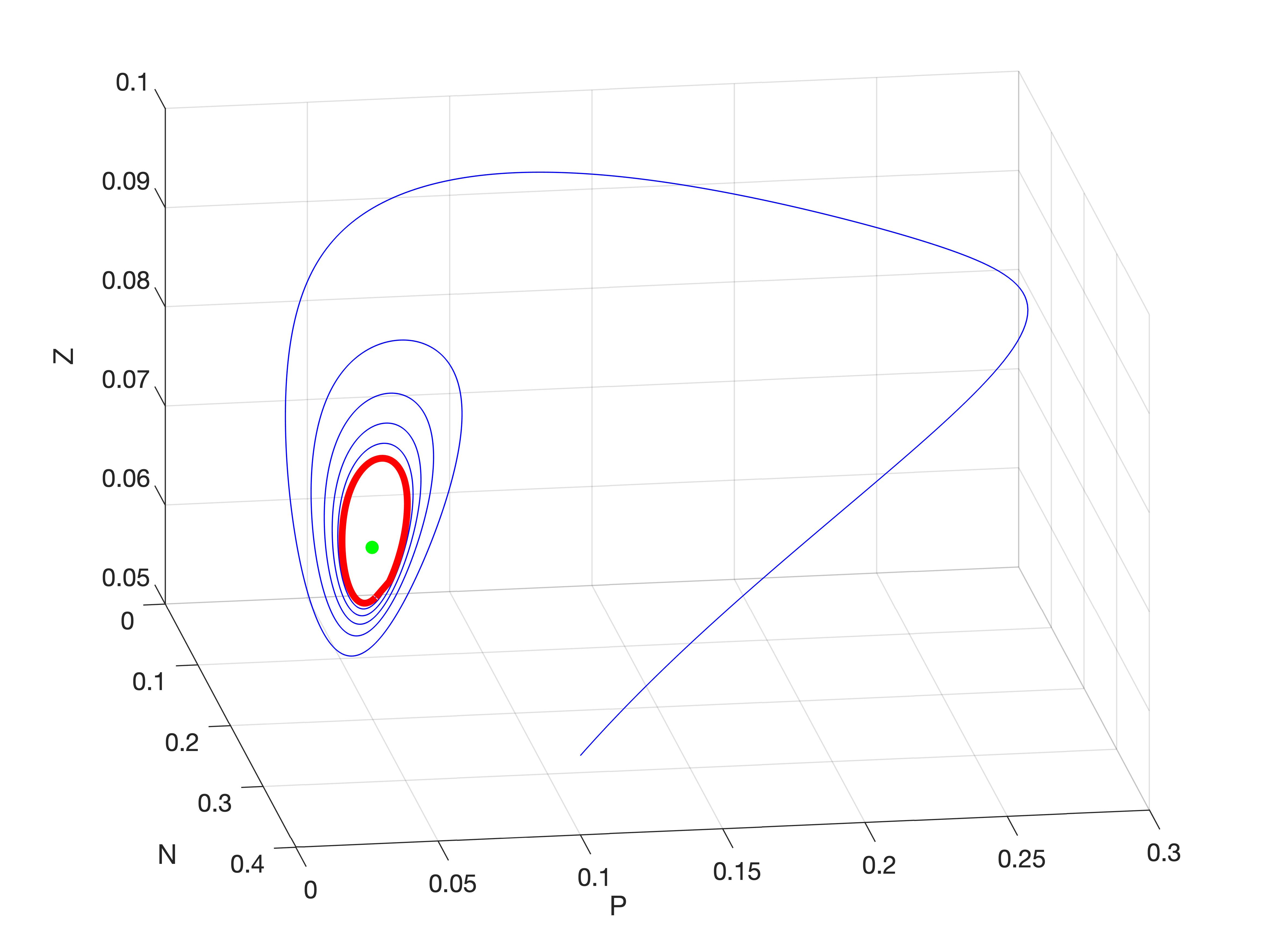}
    \caption{The phase portrait of the NPZ model.}
    \label{fig2.1}
\end{figure}
Figure \ref{fig2.1} illustrates the phase diagram of the NPZ system with the parameters set as described above. The red line represents a limit cycle, while the green dot indicates the equilibrium point of the system. It can be observed that the system, starting from a certain initial position, exhibits a coexistence of both a limit cycle and equilibrium point. Moreover, both the equilibrium point and the limit cycle are stable.

%The Jacobian matrix is given as follows:
%$\begin{pmatrix}
 %-\frac{y a e}{(b+cy)(e+x)^2}-k & -\frac{x a b}{(e+x)(b+cy)^2}+r+\frac{2 \beta \lambda z y \mu^2}{(\mu^2+y^2)^2} & \frac{\beta \lambda y^2}{\mu^2+y^2}+2 \gamma d z\\
 %\frac{e a y}{(e+x)^2(b+cy)} & \frac{x a b}{(e+x)(b+cy)^2}-r-\frac{2 \lambda y z \mu^2}{(\mu^2+y^2)^2}-s-k & -\frac{\lambda y^2}{\mu^2+y^2} \\
 %0 & \frac{2 \alpha \lambda y z \mu^2}{(\mu^2+y^2)^2} & \frac{\alpha \lambda y^2}{\mu^2+y^2}-2 d z 
%\end{pmatrix}$

\subsection{The stochastic model}
\indent

The population dynamics of plankton is influenced by a variety of stochastic factors, such as environmental changes, climatic fluctuations, and the randomness of biological interactions \cite{Upadhyay2007enviro,Abrams2000interaction}. Multiplicative noise can more effectively simulate the impact of these stochastic factors on the growth and interactions of plankton, bringing the model closer to the actual ecosystem. Therefore, we take the NPZ model with multiplicative noise into account. The corresponding stochastic model is as follows: 
\begin{equation}\label{eq_stochastic}
\begin{cases}
    \dot{x}=(-\frac{x}{e+x} \frac{a}{b+c y} y+r y+\frac{\beta \lambda y^2}{\mu^2+y^2} z+\gamma d z^2+k(N_0-x)) dt+\epsilon x dB_{t}^{1}, \\
    \dot{y}=(\frac{x}{e+x} \frac{a}{b+c y} y-r y-\frac{\lambda y^2}{\mu^2 +y^2} z-(s+k)y)dt+\epsilon y dB_{t}^{2},   \\
    \dot{z}=(\frac{\alpha \lambda y^2}{\mu^2 +y^2} z-d z^2) dt+\epsilon z dB_{t}^{3}, 
\end{cases}
\end{equation}
where $f(x,y,z)=-\frac{x}{e+x} \frac{a}{b+c y} y+r y+\frac{\beta \lambda y^2}{\mu^2+y^2} z+\gamma d z^2+k(N_0-x)$, $g(x,y,z)=\frac{x}{e+x} \frac{a}{b+c y} y-r y-\frac{\lambda y^2}{\mu^2 +y^2} z-(s+k)y$, $h(x,y,z)=\frac{\alpha \lambda y^2}{\mu^2 +y^2} z-d z^2$.

\begin{figure}
	\centering
\subfigure[$\epsilon=0.1$]{
\begin{minipage}[b]{0.4\textwidth}
\includegraphics[width=1\textwidth]{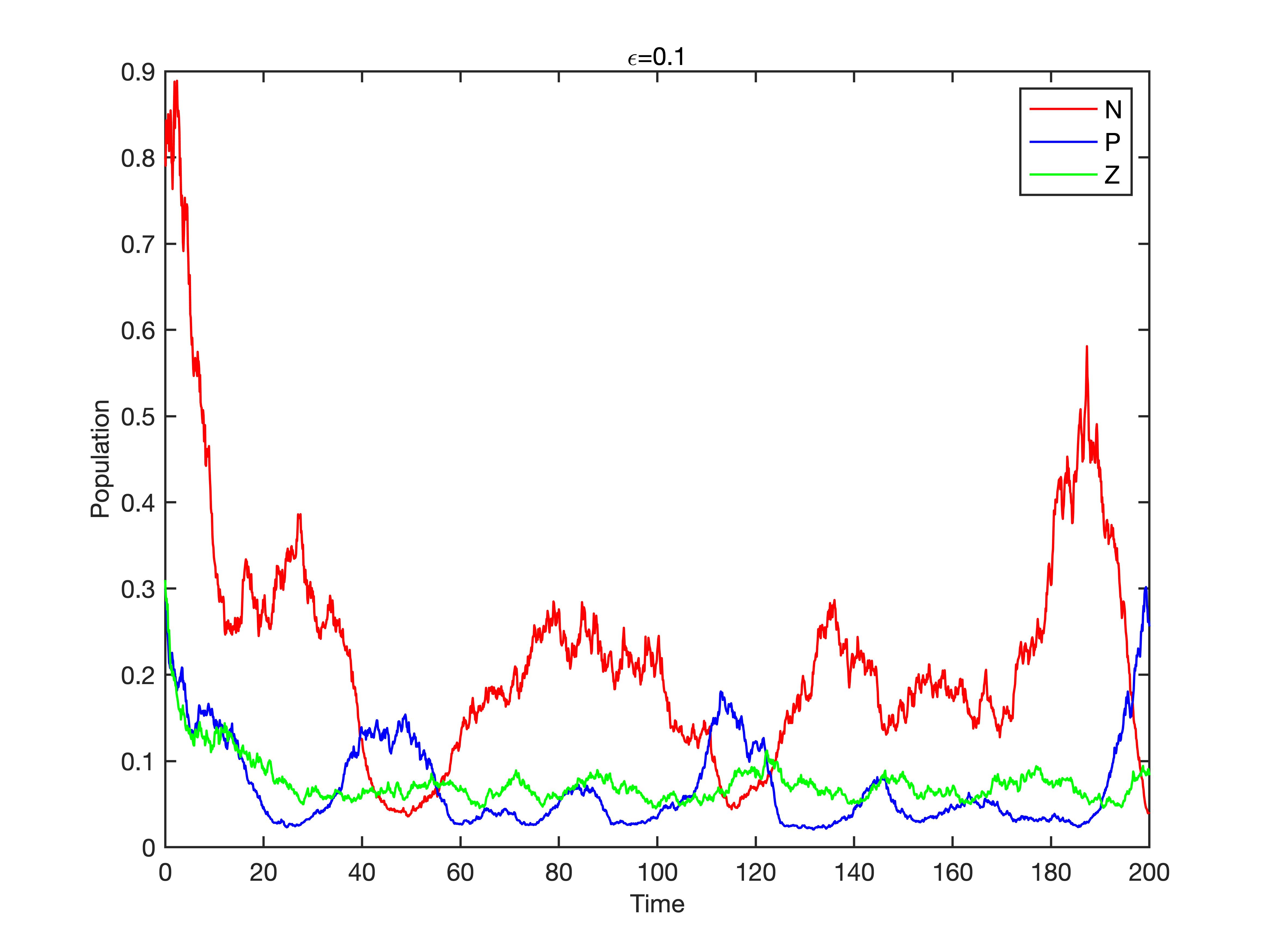} 
\end{minipage}
\label{fig:traj_0.1}
}
\subfigure[$\epsilon=0.3$]{
\begin{minipage}[b]{0.4\textwidth}
\includegraphics[width=1\textwidth]{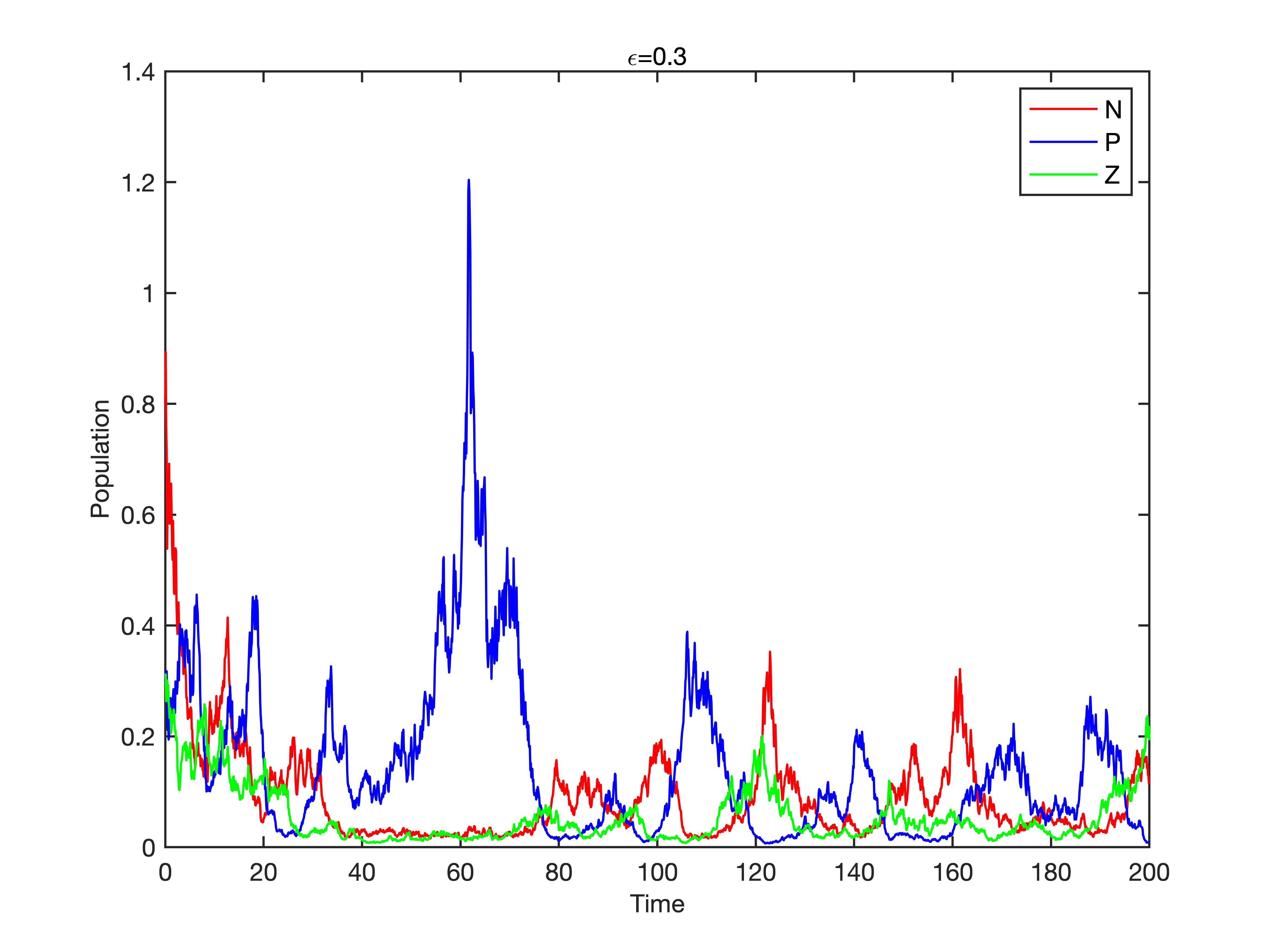}
    \end{minipage}
\label{fig:traj_0.3}
 }\\
 
 \subfigure[$\epsilon=0.5$]{
\begin{minipage}[b]{0.4\textwidth}
\includegraphics[width=1\textwidth]{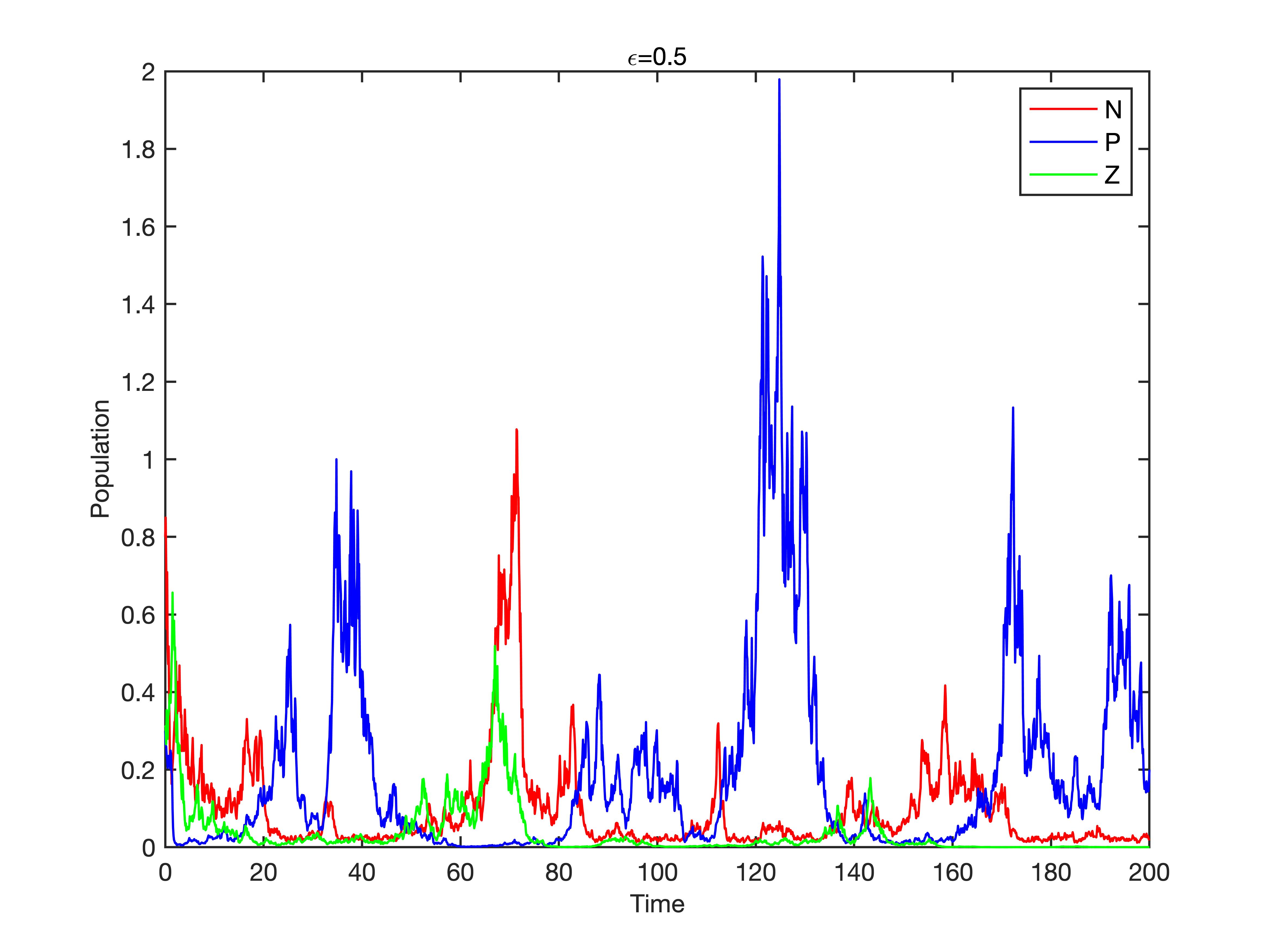}
    \end{minipage}
    
\label{fig:traj_0.5}
 }
 \subfigure[$\epsilon=0.7$]{
\begin{minipage}[b]{0.4\textwidth}
\includegraphics[width=1\textwidth]{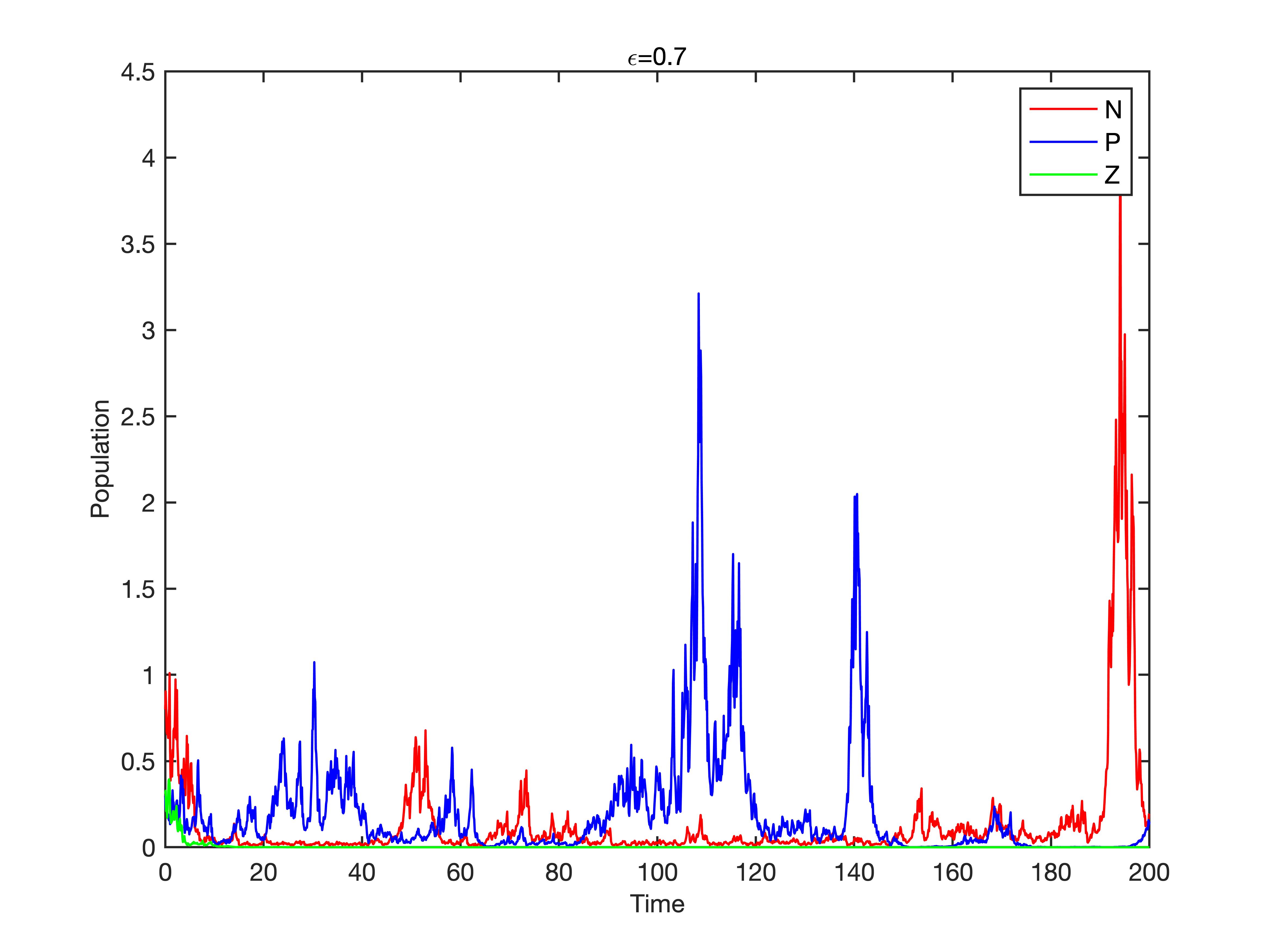}
    \end{minipage}
\label{fig:traj_0.7}
 }
 \caption{The trajectories of populations with different noise intensities. (a) $\epsilon=0.1$; (b) $\epsilon=0.3$;
(c) $\epsilon=0.5$;
(d) $\epsilon=0.7$.}
\label{fig:traj}
 \end{figure}

Figure \ref{fig:traj} illustrates the stochastic trajectories of three components under varying noise intensities. Here, $N$ represents nutrients, $P$ represents phytoplankton, and $Z$ represents zooplankton. The noise intensity has a significant impact on the fluctuations of the population. When the noise intensity is relatively low, only the nutrient component exhibits noticeable fluctuations, while the fluctuations in phytoplankton are much smaller in comparison. Zooplankton, on the other hand, remains almost unchanged. As the noise intensity increases, the amplitude of fluctuations in phytoplankton becomes more pronounced, and zooplankton begins to show slight variations. However, when the noise intensity reaches $0.7$, the fluctuations in both phytoplankton and zooplankton decrease. Moreover, the population of zooplankton drops to zero after $T=20$.

\section{Methods}
\subsection{Simulated Brownian motion}
\indent

In this model, Brownian motion is represented by a continuous Wiener process that features independent random time increments. The trajectory of a Brownian particle is determined by adding the initial position to a sequence of normally distributed random displacements \cite{Vaibhava2021SimulateBM}.
\begin{equation*}
    x_{n+1}=x_n+\Delta x_n, ~ x_0=x,
\end{equation*}
\begin{equation*}
    y_{n+1}=y_n+\Delta y_n, ~ y_0=y,
\end{equation*}
\begin{equation*}
    z_{n+1}=z_n+\Delta z_n, ~ z_0=z,
\end{equation*}
\begin{equation*}
    \Delta x_n=\sqrt{2 \epsilon x_0 dt} \gamma_{n-1},
\end{equation*}
\begin{equation*}
    \Delta y_n=\sqrt{2 \epsilon y_0 dt} \gamma_{n-1},
\end{equation*}
\begin{equation*}
    \Delta z_n=\sqrt{2 \epsilon z_0 dt} \gamma_{n-1},
\end{equation*}
\begin{equation*}
    \gamma_n \sim N(0,1), ~ n \ge 0,
\end{equation*}
where dt is the time increment. In MATLAB, the function randn is employed to compute displacements following a normal distribution. A parallel code simulation based on MATLAB is utilized to model the motion of $N$ Brownian particles until each particle escapes the limit cycle by crossing its boundary.

\subsection{The Onsager-Machlup action functional}
\indent

We consider a generalized stochastic differential equation with multiplicative Gaussian noise as follows:
$$dX(t)=\tilde{b}(X(t))dt+\sigma(X(t))dB(t),$$
where $b:R^3\to R^3$ is a regular function, $\sigma:R^3\to R^3$ is a $3\times 3$ matrix-valued function. Here, $B$ represents Brownian motion in $R^3$. In this NPZ model, $\sigma=\begin{pmatrix}
 \epsilon x &  & \\
  & \epsilon y & \\
  &  & \epsilon z
\end{pmatrix}$.

Onsager-Machlup action functional theory is a method for estimating the probability of a solution path in a specific neighborhood. The probability expression is as follows:
\begin{equation}\label{eq:prob}
    P(\left \| X-\omega \right \|_{t} \le \delta)\propto C(\delta,t)exp \left\{ -S(\omega,\dot{\omega})\right\},
\end{equation}
where $\delta$ is positive and sufficiently small, and $C(\delta,t)=P(\left \| B_t \right \|\le \delta)$ represents the probability of Brownian motion in the $\delta$ tube.

The Onsager–Machlup action functional is defined as
$$S(\omega,\dot{\omega})=\frac{1}{2}\int_{0}^{t} [(\dot{\omega}-b(\omega))V(\omega)(\dot{\omega}-b(\omega))+div b(\omega)-\frac{1}{6} R(\omega)]dt,$$
where $V(\omega)=(\sigma \sigma^T)^{-1}$, $R(\omega)$ is the scalar curvature with respect to the Riemannian metric induced by $V(\omega)$, and $b^{i}(\omega)=\tilde{b}^{i}(\omega)-\frac{1}{2} \sum_{lj}^{}(V^{-1}(\omega))^{lj}\Gamma_{lj}^{i}$ is the $i-$th component of $b$ \cite{Coulibaly-Pasquier2014b,Zeitouni1988b}. Here, $\Gamma_{lj}^{i}$ is the Christoffel symbols associated with this Riemannian
metric \cite{Jost2008RiemannianGA} and it satisfies
$$\Gamma_{l j}^{i}=\frac{1}{2} \sum_{m}^{}g^{i m}(\frac{\partial}{\partial x^j}g_{l m}+\frac{\partial}{\partial x^l}g_{j m}-\frac{\partial}{\partial x^m}g_{l j}),$$
where $g^{ij}(\omega)$ is the inverse of the Riemannian metric $g_{ij}(\omega)=V(\omega)$. And the divergence $divb(\omega)$ is defined as
$$divb(\omega)=\frac{1}{\sqrt{|V(\omega)|}}\sum_{i}^{}\frac{\partial}{\partial \omega_i}(b^i(\omega)\sqrt{|V(\omega)|}),$$
where $|V(\omega)|$ is the determinate of Riemannian metric. In our model, the Riemannian metric is $$V(x,y,z)=(g_{ij})_{3\times 3}=(\sigma \sigma^T)^{-1}=\begin{pmatrix}
 \frac{1}{\epsilon^2 x^2} &  & \\
  &\frac{1}{\epsilon^2 y^2}& \\
  &  &\frac{1}{\epsilon^2 z^2}\end{pmatrix},$$
and its inverse matrix is $$(g^{ij})_{3\times 3}=(V(x,y,z))^{-1}=\begin{pmatrix}
 \epsilon^2 x^2 &  & \\
  &\epsilon^2 y^2& \\
  &  &\epsilon^2 z^2\end{pmatrix}.$$ 
The determinate of $V$ is $|V(x,y,z)|=\frac{1}{\epsilon^6 x^2 y^2 z^2}$.

According to the expression of the Christoffel symbols, we can get the following Christoffel symbols:
\begin{equation*}
\begin{aligned}
     \Gamma_{11}^{1} &= \frac{1}{2}g^{11}(\frac{\partial}{\partial x}g_{11}+\frac{\partial}{\partial x}g_{11}-\frac{\partial}{\partial x}g_{11})+\frac{1}{2}g^{12}(\frac{\partial}{\partial x}g_{12}+\frac{\partial}{\partial x}g_{12}-\frac{\partial}{\partial y}g_{11})+\frac{1}{2}g^{13}(\frac{\partial}{\partial x}g_{13}+\frac{\partial}{\partial x}g_{13}-\frac{\partial}{\partial z}g_{11}) \\
     &=-\frac{1}{x},
\end{aligned}
\end{equation*}
\begin{equation*}
\begin{aligned}
    \Gamma_{12}^{1} &= \frac{1}{2}g^{11}(\frac{\partial}{\partial y}g_{11}+\frac{\partial}{\partial x}g_{21}-\frac{\partial}{\partial x}g_{12})+\frac{1}{2}g^{12}(\frac{\partial}{\partial y}g_{12}+\frac{\partial}{\partial x}g_{22}-\frac{\partial}{\partial y}g_{12})+\frac{1}{2}g^{13}(\frac{\partial}{\partial y}g_{13}+\frac{\partial}{\partial x}g_{23}-\frac{\partial}{\partial z}g_{12}) \\
    &=0,
\end{aligned}
\end{equation*}
\begin{equation*}
\begin{aligned}
    \Gamma_{13}^{1} &= \frac{1}{2}g^{11}(\frac{\partial}{\partial z}g_{11}+\frac{\partial}{\partial x}g_{31}-\frac{\partial}{\partial x}g_{13})+\frac{1}{2}g^{12}(\frac{\partial}{\partial z}g_{12}+\frac{\partial}{\partial x}g_{32}-\frac{\partial}{\partial y}g_{13})+\frac{1}{2}g^{13}(\frac{\partial}{\partial z}g_{13}+\frac{\partial}{\partial x}g_{33}-\frac{\partial}{\partial z}g_{13}) \\
    &=0, 
\end{aligned}
\end{equation*}
\begin{equation*}
\begin{aligned}
    \Gamma_{21}^{1} &= \frac{1}{2}g^{11}(\frac{\partial}{\partial x}g_{21}+\frac{\partial}{\partial y}g_{11}-\frac{\partial}{\partial x}g_{21})+\frac{1}{2}g^{12}(\frac{\partial}{\partial x}g_{22}+\frac{\partial}{\partial y}g_{12}-\frac{\partial}{\partial y}g_{21})+\frac{1}{2}g^{13}(\frac{\partial}{\partial x}g_{23}+\frac{\partial}{\partial y}g_{13}-\frac{\partial}{\partial z}g_{21}) \\
    &=0, 
\end{aligned}
\end{equation*}
\begin{equation*}
\begin{aligned}
    \Gamma_{22}^{1} &= \frac{1}{2}g^{11}(\frac{\partial}{\partial y}g_{21}+\frac{\partial}{\partial y}g_{21}-\frac{\partial}{\partial x}g_{22})+\frac{1}{2}g^{12}(\frac{\partial}{\partial y}g_{22}+\frac{\partial}{\partial y}g_{22}-\frac{\partial}{\partial y}g_{22})+\frac{1}{2}g^{13}(\frac{\partial}{\partial y}g_{23}+\frac{\partial}{\partial y}g_{23}-\frac{\partial}{\partial z}g_{22}) \\
    &=0,
\end{aligned}
\end{equation*}
\begin{equation*}
\begin{aligned}
    \Gamma_{23}^{1} &= \frac{1}{2}g^{11}(\frac{\partial}{\partial z}g_{21}+\frac{\partial}{\partial y}g_{31}-\frac{\partial}{\partial x}g_{23})+\frac{1}{2}g^{12}(\frac{\partial}{\partial z}g_{22}+\frac{\partial}{\partial y}g_{32}-\frac{\partial}{\partial y}g_{23})+\frac{1}{2}g^{13}(\frac{\partial}{\partial z}g_{23}+\frac{\partial}{\partial y}g_{33}-\frac{\partial}{\partial z}g_{23}) \\
    &=0,
\end{aligned}
\end{equation*}
\begin{equation*}
\begin{aligned}
    \Gamma_{31}^{1} &= \frac{1}{2}g^{11}(\frac{\partial}{\partial x}g_{31}+\frac{\partial}{\partial z}g_{11}-\frac{\partial}{\partial x}g_{31})+\frac{1}{2}g^{12}(\frac{\partial}{\partial x}g_{32}+\frac{\partial}{\partial z}g_{12}-\frac{\partial}{\partial y}g_{31})+\frac{1}{2}g^{13}(\frac{\partial}{\partial x}g_{33}+\frac{\partial}{\partial z}g_{13}-\frac{\partial}{\partial z}g_{31}) \\
    &=0,
\end{aligned}
\end{equation*}
\begin{equation*}
\begin{aligned}
    \Gamma_{32}^{1} &= \frac{1}{2}g^{11}(\frac{\partial}{\partial y}g_{31}+\frac{\partial}{\partial z}g_{21}-\frac{\partial}{\partial x}g_{32})+\frac{1}{2}g^{12}(\frac{\partial}{\partial y}g_{32}+\frac{\partial}{\partial z}g_{22}-\frac{\partial}{\partial y}g_{32})+\frac{1}{2}g^{13}(\frac{\partial}{\partial y}g_{33}+\frac{\partial}{\partial z}g_{23}-\frac{\partial}{\partial z}g_{32}) \\
    &=0,
\end{aligned}
\end{equation*}
\begin{equation*}
\begin{aligned}
    \Gamma_{33}^{1} &= \frac{1}{2}g^{11}(\frac{\partial}{\partial z}g_{31}+\frac{\partial}{\partial z}g_{31}-\frac{\partial}{\partial x}g_{33})+\frac{1}{2}g^{12}(\frac{\partial}{\partial z}g_{32}+\frac{\partial}{\partial z}g_{32}-\frac{\partial}{\partial y}g_{33})+\frac{1}{2}g^{13}(\frac{\partial}{\partial z}g_{33}+\frac{\partial}{\partial z}g_{33}-\frac{\partial}{\partial z}g_{33}) \\
    &=0,
\end{aligned}
\end{equation*}
\begin{equation*}
\begin{aligned}
    \Gamma_{11}^{2} &= \frac{1}{2}g^{21}(\frac{\partial}{\partial x}g_{11}+\frac{\partial}{\partial x}g_{11}-\frac{\partial}{\partial x}g_{11})+\frac{1}{2}g^{22}(\frac{\partial}{\partial x}g_{12}+\frac{\partial}{\partial x}g_{12}-\frac{\partial}{\partial y}g_{11})+\frac{1}{2}g^{23}(\frac{\partial}{\partial x}g_{13}+\frac{\partial}{\partial x}g_{13}-\frac{\partial}{\partial z}g_{11}) \\
    &=0,
\end{aligned}
\end{equation*}
\begin{equation*}
\begin{aligned}
    \Gamma_{12}^{2} &= \frac{1}{2}g^{21}(\frac{\partial}{\partial y}g_{11}+\frac{\partial}{\partial x}g_{21}-\frac{\partial}{\partial x}g_{12})+\frac{1}{2}g^{22}(\frac{\partial}{\partial y}g_{12}+\frac{\partial}{\partial x}g_{22}-\frac{\partial}{\partial y}g_{12})+\frac{1}{2}g^{23}(\frac{\partial}{\partial y}g_{13}+\frac{\partial}{\partial x}g_{23}-\frac{\partial}{\partial z}g_{12}) \\
    &=0,
\end{aligned}
\end{equation*}
\begin{equation*}
\begin{aligned}
    \Gamma_{13}^{2} &= \frac{1}{2}g^{21}(\frac{\partial}{\partial z}g_{12}+\frac{\partial}{\partial x}g_{31}-\frac{\partial}{\partial x}g_{13})+\frac{1}{2}g^{22}(\frac{\partial}{\partial z}g_{12}+\frac{\partial}{\partial x}g_{32}-\frac{\partial}{\partial y}g_{13})+\frac{1}{2}g^{23}(\frac{\partial}{\partial z}g_{13}+\frac{\partial}{\partial x}g_{33}-\frac{\partial}{\partial z}g_{13}) \\
    &=0,
\end{aligned}
\end{equation*}
\begin{equation*}
\begin{aligned}
    \Gamma_{21}^{2} &= \frac{1}{2}g^{21}(\frac{\partial}{\partial x}g_{21}+\frac{\partial}{\partial y}g_{11}-\frac{\partial}{\partial x}g_{21})+\frac{1}{2}g^{22}(\frac{\partial}{\partial x}g_{22}+\frac{\partial}{\partial y}g_{12}-\frac{\partial}{\partial y}g_{21})+\frac{1}{2}g^{23}(\frac{\partial}{\partial x}g_{23}+\frac{\partial}{\partial y}g_{13}-\frac{\partial}{\partial z}g_{21}) \\
    &=0,
\end{aligned}
\end{equation*}
\begin{equation*}
\begin{aligned}
    \Gamma_{22}^{2} &= \frac{1}{2}g^{21}(\frac{\partial}{\partial y}g_{21}+\frac{\partial}{\partial y}g_{21}-\frac{\partial}{\partial x}g_{22})+\frac{1}{2}g^{22}(\frac{\partial}{\partial y}g_{22}+\frac{\partial}{\partial y}g_{22}-\frac{\partial}{\partial y}g_{22})+\frac{1}{2}g^{23}(\frac{\partial}{\partial y}g_{23}+\frac{\partial}{\partial y}g_{23}-\frac{\partial}{\partial z}g_{22}) \\
    &=-\frac{1}{y},
\end{aligned}
\end{equation*}
\begin{equation*}
\begin{aligned}
    \Gamma_{23}^{2} &= \frac{1}{2}g^{21}(\frac{\partial}{\partial z}g_{21}+\frac{\partial}{\partial y}g_{31}-\frac{\partial}{\partial x}g_{23})+\frac{1}{2}g^{22}(\frac{\partial}{\partial z}g_{22}+\frac{\partial}{\partial y}g_{32}-\frac{\partial}{\partial y}g_{23})+\frac{1}{2}g^{23}(\frac{\partial}{\partial z}g_{23}+\frac{\partial}{\partial y}g_{33}-\frac{\partial}{\partial z}g_{23}) \\
    &=0,
\end{aligned}
\end{equation*}
\begin{equation*}
\begin{aligned}
    \Gamma_{31}^{2} &= \frac{1}{2}g^{21}(\frac{\partial}{\partial x}g_{31}+\frac{\partial}{\partial z}g_{11}-\frac{\partial}{\partial x}g_{31})+\frac{1}{2}g^{22}(\frac{\partial}{\partial x}g_{32}+\frac{\partial}{\partial z}g_{12}-\frac{\partial}{\partial y}g_{31})+\frac{1}{2}g^{23}(\frac{\partial}{\partial x}g_{33}+\frac{\partial}{\partial z}g_{13}-\frac{\partial}{\partial z}g_{31}) \\
    &=0,
\end{aligned}
\end{equation*}
\begin{equation*}
\begin{aligned}
    \Gamma_{32}^{2} &= \frac{1}{2}g^{21}(\frac{\partial}{\partial y}g_{31}+\frac{\partial}{\partial z}g_{21}-\frac{\partial}{\partial x}g_{32})+\frac{1}{2}g^{22}(\frac{\partial}{\partial y}g_{32}+\frac{\partial}{\partial z}g_{22}-\frac{\partial}{\partial y}g_{32})+\frac{1}{2}g^{23}(\frac{\partial}{\partial y}g_{33}+\frac{\partial}{\partial z}g_{23}-\frac{\partial}{\partial z}g_{32}) \\
    &=0,
\end{aligned}
\end{equation*}
\begin{equation*}
\begin{aligned}
    \Gamma_{33}^{2} &= \frac{1}{2}g^{21}(\frac{\partial}{\partial z}g_{31}+\frac{\partial}{\partial z}g_{31}-\frac{\partial}{\partial x}g_{33})+\frac{1}{2}g^{22}(\frac{\partial}{\partial z}g_{32}+\frac{\partial}{\partial z}g_{32}-\frac{\partial}{\partial y}g_{33})+\frac{1}{2}g^{23}(\frac{\partial}{\partial z}g_{33}+\frac{\partial}{\partial z}g_{33}-\frac{\partial}{\partial z}g_{33}) \\
    &=0,
\end{aligned}
\end{equation*}
\begin{equation*}
\begin{aligned}
    \Gamma_{11}^{3} &= \frac{1}{2}g^{31}(\frac{\partial}{\partial x}g_{11}+\frac{\partial}{\partial x}g_{11}-\frac{\partial}{\partial x}g_{11})+\frac{1}{2}g^{32}(\frac{\partial}{\partial x}g_{12}+\frac{\partial}{\partial x}g_{12}-\frac{\partial}{\partial y}g_{11})+\frac{1}{2}g^{33}(\frac{\partial}{\partial x}g_{13}+\frac{\partial}{\partial x}g_{13}-\frac{\partial}{\partial z}g_{11}) \\
    &=0,
\end{aligned}
\end{equation*}
\begin{equation*}
\begin{aligned}
    \Gamma_{12}^{3} &= \frac{1}{2}g^{31}(\frac{\partial}{\partial y}g_{11}+\frac{\partial}{\partial x}g_{21}-\frac{\partial}{\partial x}g_{12})+\frac{1}{2}g^{32}(\frac{\partial}{\partial y}g_{12}+\frac{\partial}{\partial x}g_{22}-\frac{\partial}{\partial y}g_{12})+\frac{1}{2}g^{33}(\frac{\partial}{\partial y}g_{13}+\frac{\partial}{\partial x}g_{23}-\frac{\partial}{\partial z}g_{12}) \\
    &=0,
\end{aligned}
\end{equation*}
\begin{equation*}
\begin{aligned}
    \Gamma_{13}^{3} &= \frac{1}{2}g^{31}(\frac{\partial}{\partial z}g_{11}+\frac{\partial}{\partial x}g_{31}-\frac{\partial}{\partial x}g_{13})+\frac{1}{2}g^{32}(\frac{\partial}{\partial z}g_{12}+\frac{\partial}{\partial x}g_{32}-\frac{\partial}{\partial y}g_{13})+\frac{1}{2}g^{33}(\frac{\partial}{\partial z}g_{13}+\frac{\partial}{\partial x}g_{33}-\frac{\partial}{\partial z}g_{13}) \\
    &=0,
\end{aligned}
\end{equation*}
\begin{equation*}
\begin{aligned}
    \Gamma_{21}^{3} &= \frac{1}{2}g^{31}(\frac{\partial}{\partial x}g_{21}+\frac{\partial}{\partial y}g_{11}-\frac{\partial}{\partial x}g_{21})+\frac{1}{2}g^{32}(\frac{\partial}{\partial x}g_{22}+\frac{\partial}{\partial y}g_{12}-\frac{\partial}{\partial y}g_{21})+\frac{1}{2}g^{33}(\frac{\partial}{\partial x}g_{23}+\frac{\partial}{\partial y}g_{13}-\frac{\partial}{\partial z}g_{21}) \\
    &=0,
\end{aligned}
\end{equation*}
\begin{equation*}
\begin{aligned}
    \Gamma_{22}^{3} &= \frac{1}{2}g^{31}(\frac{\partial}{\partial y}g_{21}+\frac{\partial}{\partial y}g_{21}-\frac{\partial}{\partial x}g_{22})+\frac{1}{2}g^{32}(\frac{\partial}{\partial y}g_{22}+\frac{\partial}{\partial y}g_{22}-\frac{\partial}{\partial y}g_{22})+\frac{1}{2}g^{33}(\frac{\partial}{\partial y}g_{23}+\frac{\partial}{\partial y}g_{23}-\frac{\partial}{\partial z}g_{22}) \\
    &=0,
\end{aligned}
\end{equation*}
\begin{equation*}
\begin{aligned}
    \Gamma_{23}^{3} &= \frac{1}{2}g^{31}(\frac{\partial}{\partial z}g_{21}+\frac{\partial}{\partial y}g_{31}-\frac{\partial}{\partial x}g_{23})+\frac{1}{2}g^{32}(\frac{\partial}{\partial z}g_{22}+\frac{\partial}{\partial y}g_{32}-\frac{\partial}{\partial y}g_{23})+\frac{1}{2}g^{33}(\frac{\partial}{\partial z}g_{23}+\frac{\partial}{\partial y}g_{33}-\frac{\partial}{\partial z}g_{23}) \\
    &=0,
\end{aligned}
\end{equation*}
\begin{equation*}
\begin{aligned}
    \Gamma_{31}^{3} &= \frac{1}{2}g^{31}(\frac{\partial}{\partial x}g_{31}+\frac{\partial}{\partial z}g_{11}-\frac{\partial}{\partial x}g_{31})+\frac{1}{2}g^{32}(\frac{\partial}{\partial x}g_{32}+\frac{\partial}{\partial z}g_{12}-\frac{\partial}{\partial y}g_{31})+\frac{1}{2}g^{33}(\frac{\partial}{\partial x}g_{33}+\frac{\partial}{\partial z}g_{13}-\frac{\partial}{\partial z}g_{31}) \\
    &=0,
\end{aligned}
\end{equation*}
\begin{equation*}
\begin{aligned}
    \Gamma_{32}^{3} &= \frac{1}{2}g^{31}(\frac{\partial}{\partial y}g_{31}+\frac{\partial}{\partial z}g_{21}-\frac{\partial}{\partial x}g_{32})+\frac{1}{2}g^{32}(\frac{\partial}{\partial y}g_{32}+\frac{\partial}{\partial z}g_{22}-\frac{\partial}{\partial y}g_{32})+\frac{1}{2}g^{33}(\frac{\partial}{\partial y}g_{33}+\frac{\partial}{\partial z}g_{23}-\frac{\partial}{\partial z}g_{32}) \\
    &=0,
\end{aligned}
\end{equation*}
\begin{equation*}
\begin{aligned}
    \Gamma_{33}^{3} &= \frac{1}{2}g^{31}(\frac{\partial}{\partial z}g_{31}+\frac{\partial}{\partial z}g_{31}-\frac{\partial}{\partial x}g_{33})+\frac{1}{2}g^{32}(\frac{\partial}{\partial z}g_{32}+\frac{\partial}{\partial z}g_{32}-\frac{\partial}{\partial y}g_{33})+\frac{1}{2}g^{33}(\frac{\partial}{\partial z}g_{33}+\frac{\partial}{\partial z}g_{33}-\frac{\partial}{\partial z}g_{33}) \\
    &=-\frac{1}{z}.
\end{aligned}
\end{equation*}
The modified field is 
\begin{equation*}
    b^{1}(x,y,z)=f(x,y,z)-\frac{1}{2} (\epsilon^2 x^2) \frac{-1}{x}=f(x,y,z)+\frac{1}{2}\epsilon^2 x,
\end{equation*}
\begin{equation*}
    b^{2}(x,y,z)=g(x,y,z)-\frac{1}{2} (\epsilon^2 y^2) \frac{-1}{y}=g(x,y,z)+\frac{1}{2}\epsilon^2 y,
\end{equation*}
\begin{equation*}
    b^{3}(x,y,z)=h(x,y,z)-\frac{1}{2} (\epsilon^2 z^2) \frac{-1}{z}=h(x,y,z)+\frac{1}{2}\epsilon^2 z.
\end{equation*}
And the divergence $divb(\omega)$ is computed as
\begin{align*}
    divb(x,y,z) &=\frac{1}{\sqrt{|V(x,y,z)|}}\sum_{i}^{}\frac{\partial}{\partial \omega_i}(b^i(\omega)\sqrt{|V(x,y,z)|}) \\
    &=f_{x}+g_{y}+h_{z}-\frac{f}{x}-\frac{g}{y}-\frac{h}{z}  \\
    &=\frac{1}{e+x}\frac{a}{b+cy} y (1-e)-\frac{1}{x}(r y+\frac{\beta \lambda y^2}{\mu^2+y^2}z+\gamma d z^2+k N_0)+\frac{x}{e+x}\frac{a}{b+cy}(b-1) \\
    &+\frac{\lambda y}{\mu^2+y^2}(1-\frac{2 \mu^2}{\mu^2+y^2}z-d z).
\end{align*}

The scalar curvature is defined as
the trace of the Ricci curvature tensor with respect to the Riemannian metric $$R=g^{ij}R_{ij},$$ 
where $R_{ij}$ is the Ricci curvature tensor. By \cite{Jost2008RiemannianGA}, we have 
$$R=g^{ik}R_{ikm}^{m}=g^{ik}(\frac{\partial}{\partial x^k} \Gamma_{im}^{m}-\frac{\partial}{\partial x^m} \Gamma_{ki}^{m}+ \Gamma_{ka}^{m} \Gamma_{mi}^{a}-\Gamma_{ma}^{m} \Gamma_{ki}^{a}).$$
We can get the scalar curvature of model as follows:
\begin{align*}
    R=&g^{11}(\frac{\partial}{\partial x} \Gamma_{1m}^{m}-\frac{\partial}{\partial x^m} \Gamma_{11}^{m}+ \Gamma_{1a}^{m}\Gamma_{m1}^{a}-\Gamma_{ma}^{m}\Gamma_{11}^{a})+g^{22}(\frac{\partial}{\partial y} \Gamma_{2m}^{m}-\frac{\partial}{\partial x^m} \Gamma_{22}^{m}+ \Gamma_{2a}^{m}\Gamma_{m2}^{a}-\Gamma_{ma}^{m}\Gamma_{22}^{a})\\
    &+g^{33}(\frac{\partial}{\partial z} \Gamma_{3m}^{m}-\frac{\partial}{\partial x^m} \Gamma_{33}^{m}+ \Gamma_{3a}^{m}\Gamma_{m3}^{a}-\Gamma_{ma}^{m}\Gamma_{33}^{a})\\
    =&g^{11}(\frac{\partial}{\partial x} \Gamma_{11}^{1}-\frac{\partial}{\partial x} \Gamma_{11}^{1}+ \Gamma_{11}^{1}\Gamma_{11}^{1}-\Gamma_{11}^{1}\Gamma_{11}^{1})+g^{22}(\frac{\partial}{\partial y} \Gamma_{22}^{2}-\frac{\partial}{\partial y} \Gamma_{22}^{2}+ \Gamma_{22}^{2}\Gamma_{22}^{2}-\Gamma_{22}^{2}\Gamma_{22}^{2})\\
    &+g^{33}(\frac{\partial}{\partial z} \Gamma_{33}^{3}-\frac{\partial}{\partial z} \Gamma_{33}^{3}+ \Gamma_{33}^{3}\Gamma_{33}^{3}-\Gamma_{33}^{3}\Gamma_{33}^{3})\\
    =&0.
\end{align*}
Then we obtain the following expression of Lagrangian:
\begin{align*}
    L(\omega,\dot{\omega}) =&\frac{1}{2}[(\dot{\omega}-b^i(x,y,z))V(x,y,z)(\dot{\omega}-b^i(x,y,z))^T+divb(x,y,z)-\frac{1}{6}R] \\
    =&\frac{1}{2}\frac{1}{\epsilon^2 x^2}(\dot{x}-f(x,y,z)-\frac{1}{2}\epsilon^2 x)^2+\frac{1}{2}\frac{1}{\epsilon^2 y^2}(\dot{y}-g(x,y,z)-\frac{1}{2}\epsilon^2 y)^2+\frac{1}{2}\frac{1}{\epsilon^2 z^2}(\dot{z}-h(x,y,z)-\frac{1}{2}\epsilon^2 z)^2 \\
    &+\frac{1}{2}(f_{x}+g_{y}+h_{z}-\frac{f}{x}-\frac{g}{y}-\frac{h}{z}) \\
    =&\frac{1}{2}\frac{1}{\epsilon^2 x^2}(\dot{x}+\frac{x}{e+x} \frac{a}{b+cy} y-r y-\frac{\beta \lambda y^2}{\mu^2+y^2}z-\gamma d z^2-k(N_0-x)-\frac{1}{2}\epsilon^2 x)^2+\frac{1}{2}\frac{1}{\epsilon^2 y^2}(\dot{y} \\
    &-\frac{x}{e+x}\frac{a}{b+cy}y+ry+\frac{\lambda y^2}{\mu^2+y^2}z+(s+k)y-\frac{1}{2}\epsilon^2 y)^2+\frac{1}{2}\frac{1}{\epsilon^2 z^2}(\dot{z}-\frac{\alpha \lambda y^2}{\mu^2+y^2}z+d z^2-\frac{1}{2}\epsilon^2 z)^2 \\
    &+\frac{1}{2}(\frac{1}{e+x}\frac{a}{b+cy} y (1-e)-\frac{1}{x}(r y+\frac{\beta \lambda y^2}{\mu^2+y^2}z+\gamma d z^2+k N_0)+\frac{x}{e+x}\frac{a}{b+cy}(b-1) \\
    &+\frac{\lambda y}{\mu^2+y^2}(1-\frac{2 \mu^2}{\mu^2+y^2}z-d z)).
\end{align*}

The minimum value of the Lagrangian means the largest probability of the solution paths on a small tube. The most probable transition pathway between the two metastable states satisfies
the Euler-Lagrange equation
$$\frac{\mathrm{d}}{\mathrm{d}t} \frac{\partial}{\partial \dot{\omega}}L(\omega,\dot{\omega})=\frac{\partial}{\partial \omega}L(\omega,\dot{\omega}).$$

We use the Euler-Lagrange equation to obtain the following ODEs for $x$, $y$ and $z$:
\begin{equation}
    \begin{cases}
    \begin{aligned}
        \ddot{x} =&\frac{1}{x}\dot{x}^2+f_{y}(x,y,z)\dot{y}+f_{z}(x,y,z)\dot{z}-\frac{1}{x}f^2(x,y,z)+f(x,y,z) f_x(x,y,z) \\
        &-\frac{x^2}{y^2}(\dot{y}-g(x,y,z))g_x(x,y,z)
        -\frac{x^2}{z^2}(\dot{z}-h(x,y,z))h_x(x,y,z) \\
        &+\frac{\epsilon^2 x^2}{2}(f_{xx}(x,y,z)+g_{yx}(x,y,z)+h_{zx}(x,y,z)), \\
        \ddot{y} =&\frac{1}{y}\dot{y}^2+g_{x}(x,y,z)\dot{x}+g_{z}(x,y,z)\dot{z}-\frac{1}{y}g^2(x,y,z)+g(x,y,z) g_y(x,y,z) \\
        &-\frac{y^2}{x^2}(\dot{x}-f(x,y,z))f_y(x,y,z)
        -\frac{y^2}{z^2}(\dot{z}-h(x,y,z))h_y(x,y,z) \\
        &+\frac{\epsilon^2 y^2}{2}(f_{xy}(x,y,z)+g_{yy}(x,y,z)+h_{zy}(x,y,z)), \\
        \ddot{z} =&\frac{1}{z}\dot{z}^2+h_{x}(x,y,z)\dot{x}+h_{y}(x,y,z)\dot{y}-\frac{1}{z}h^2(x,y,z)+h(x,y,z) h_z(x,y,z) \\
        &-\frac{z^2}{x^2}(\dot{x}-f(x,y,z))f_z(x,y,z)
        -\frac{z^2}{y^2}(\dot{y}-g(x,y,z))g_z(x,y,z) \\
        &+\frac{\epsilon^2 z^2}{2}(f_{xz}(x,y,z)+g_{yz}(x,y,z)+h_{zz}(x,y,z)).
    \end{aligned}
    \end{cases}
\end{equation}

Then we transform the three-order ODEs to six first-order differential equations
as follows:
\begin{equation}
    \begin{cases}
        \begin{aligned}
        \dot{x}=&u,\\
        \dot{y}=&v,\\
        \dot{z}=&w,\\
        \dot{u}=&\frac{1}{x}u^2+f_{y}(x,y,z)v+f_{z}(x,y,z)w-\frac{1}{x}f^2(x,y,z)+f(x,y,z)f_x(x,y,z)\\
        &-\frac{x^2}{y^2}(v-g(x,y,z))g_x(x,y,z)-\frac{x^2}{z^2}(w-h(x,y,z))h_x(x,y,z) \\
        &+\frac{\epsilon^2 x^2}{2}(f_{xx}(x,y,z)+g_{yx}(x,y,z)+h_{zx}(x,y,z)),\\
        \dot{v}=&\frac{1}{y}v^2+g_{x}(x,y,z) u+g_{z}(x,y,z) w-\frac{1}{y}g^2(x,y,z)+g(x,y,z) g_y(x,y,z) \\
        &-\frac{y^2}{x^2}(u-f(x,y,z))f_y(x,y,z)
        -\frac{y^2}{z^2}(w-h(x,y,z))h_y(x,y,z) \\
        &+\frac{\epsilon^2 y^2}{2}(f_{xy}(x,y,z)+g_{yy}(x,y,z)+h_{zy}(x,y,z)), \\
        \dot{w} =&\frac{1}{z} w^2+h_{x}(x,y,z) u+h_{y}(x,y,z) v-\frac{1}{z}h^2(x,y,z)+h(x,y,z) h_z(x,y,z) \\
        &-\frac{z^2}{x^2}(u-f(x,y,z))f_z(x,y,z)
        -\frac{z^2}{y^2}(v-g(x,y,z))g_z(x,y,z) \\
        &+\frac{\epsilon^2 z^2}{2}(f_{xz}(x,y,z)+g_{yz}(x,y,z)+h_{zz}(x,y,z)).
         \end{aligned}
    \end{cases}
\end{equation}

\subsection{Neural shooting method}
\indent

The six-dimensional Cauchy boundary value problem is solved by using the neural shooting method. The main ideas are as follows:

$(1)$ Firstly, a sample dataset distributed on the limit cycle is generated. The dataset includes initial velocity and endpoint information, with a noise coefficient  set, as shown in Figure \ref{fig:end}.

$(2)$ Subsequently, the neural shooting method is employed for computation. The essence of the neural shooting method is to establish a mapping from initial velocity to endpoint. To approximate this mapping, a neural network with a structure of “$2$-$8$-$16$-$8$-$2$” is utilized. In the dataset generated in the first step, the initial velocity serves as the input to the neural network, while the points distributed around the limit cycle act as the output. The training error of the neural network is set to be less than $0.01$ to ensure the accuracy of the model.

$(3)$ The best model obtained in the second step is used to calculate the transition pathway from the equilibrium point to the limit cycle. Based on the Onsager-Machlup action functional theory, each transition pathway corresponds to a action functional value. In this way, the most probable transition pathway can be determined.

\newpage
\section{Main results}
\indent

\subsection{The most probable transition time}
\indent

\begin{figure}[H] 
\centering
\subfigure[$\epsilon=0.1$]{
    \begin{minipage}[b]{0.4\textwidth}
    \includegraphics[width=1\textwidth]{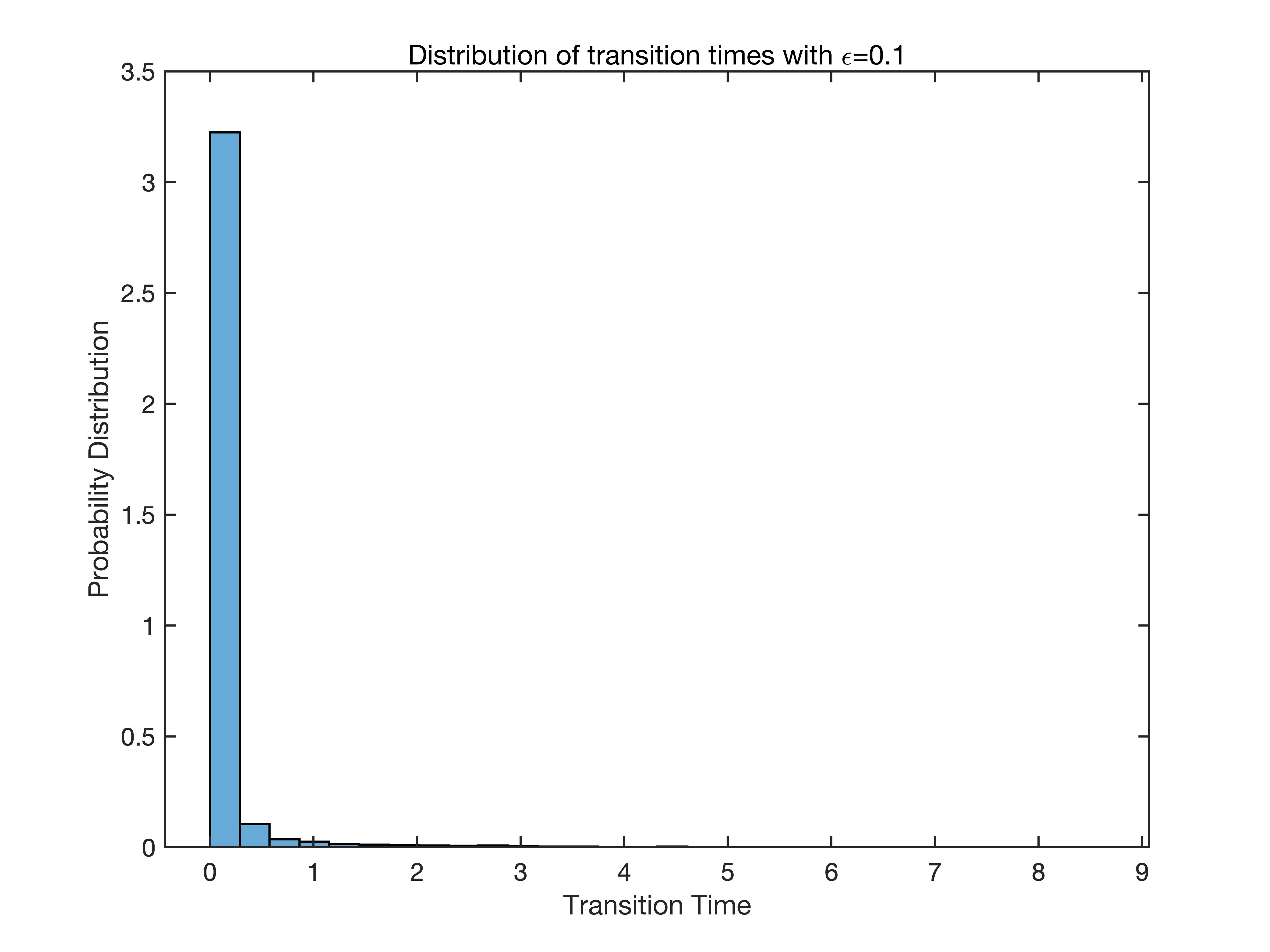} 
    \end{minipage}
    \label{fig:es0.1}
}
\subfigure[$\epsilon=0.3$]{
    \begin{minipage}[b]{0.4\textwidth}
    \includegraphics[width=1\textwidth]{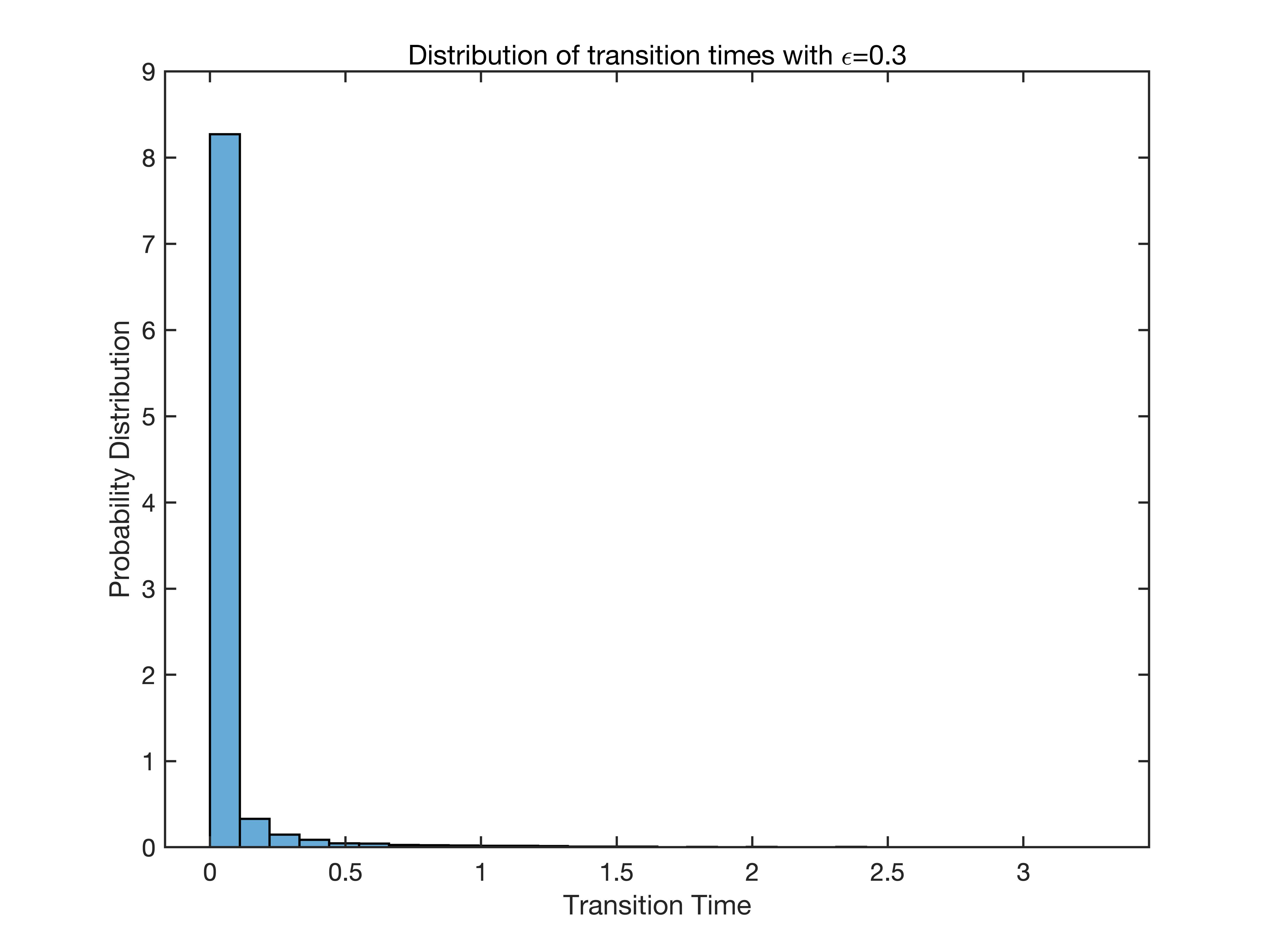}
    \end{minipage}
    \label{fig:es0.3}
}\\
 
\subfigure[$\epsilon=0.5$]{
    \begin{minipage}[b]{0.4\textwidth}
    \includegraphics[width=1\textwidth]{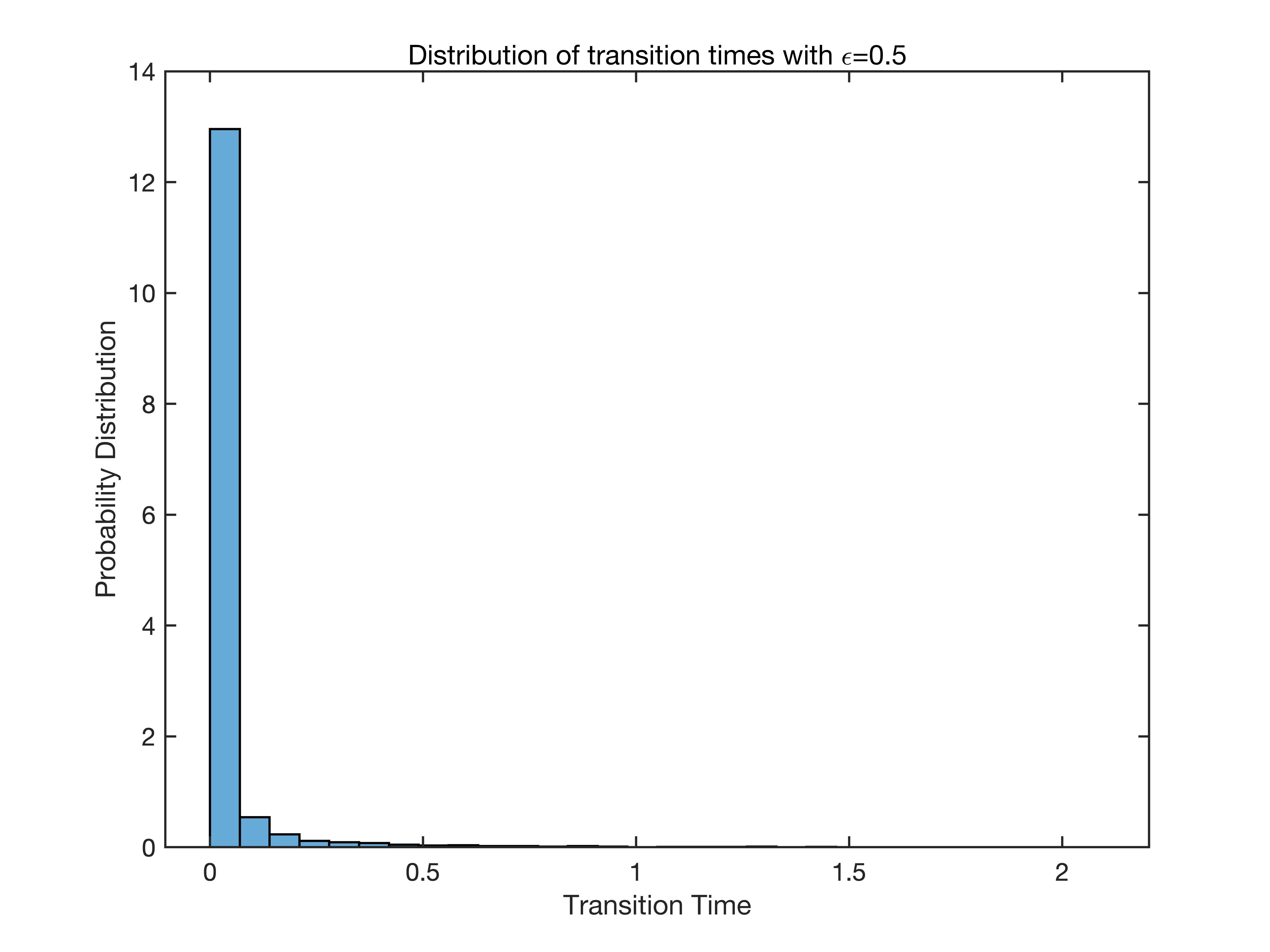}
    \end{minipage}
    \label{fig:es0.5}
}
\subfigure[$\epsilon=0.7$]{
    \begin{minipage}[b]{0.4\textwidth}
    \includegraphics[width=1\textwidth]{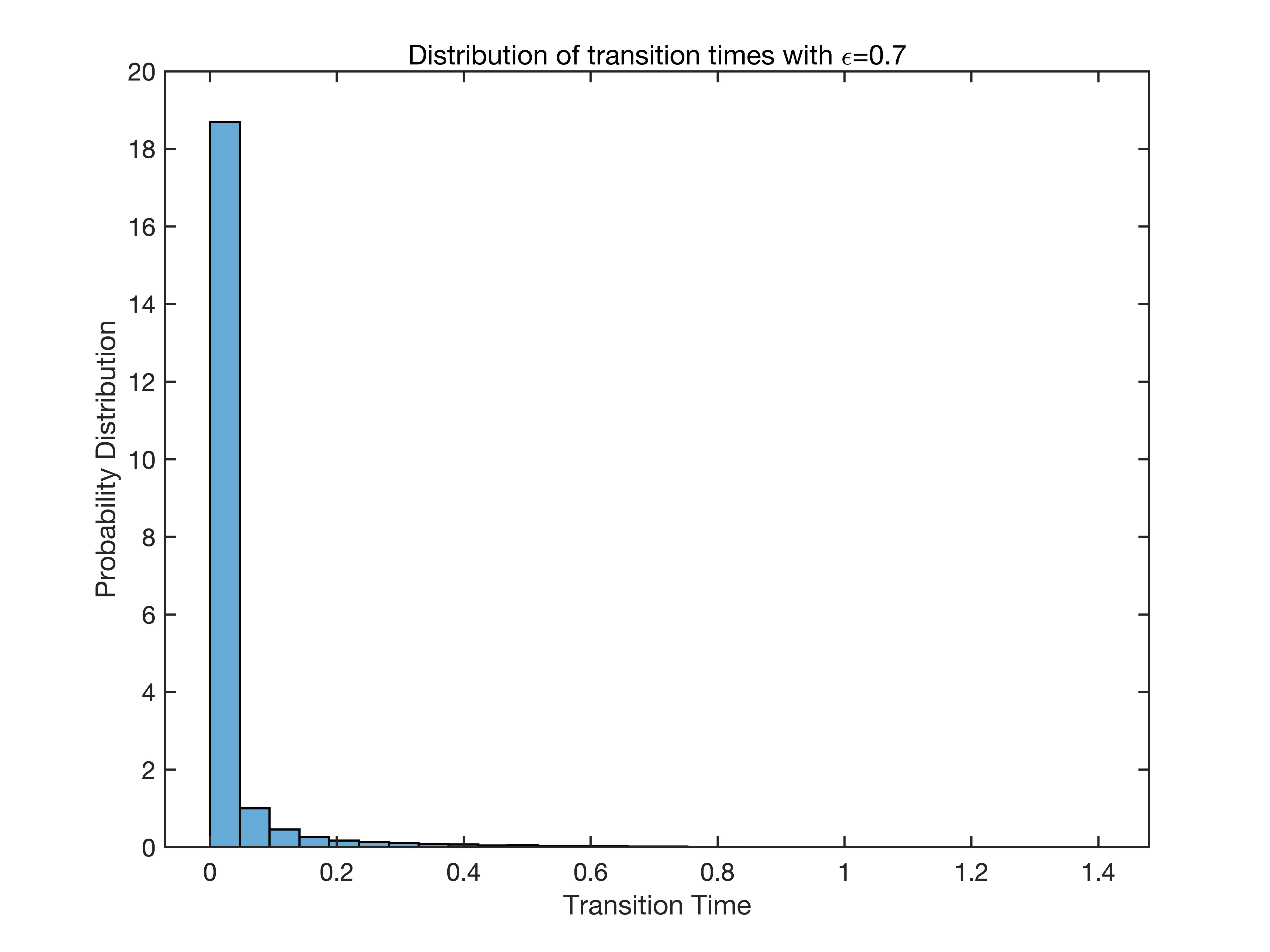}
    \end{minipage}
    \label{fig:es0.7}
}\\
\subfigure[$\epsilon=0.8$]{
    \begin{minipage}[b]{0.4\textwidth}
    \includegraphics[width=1\textwidth]{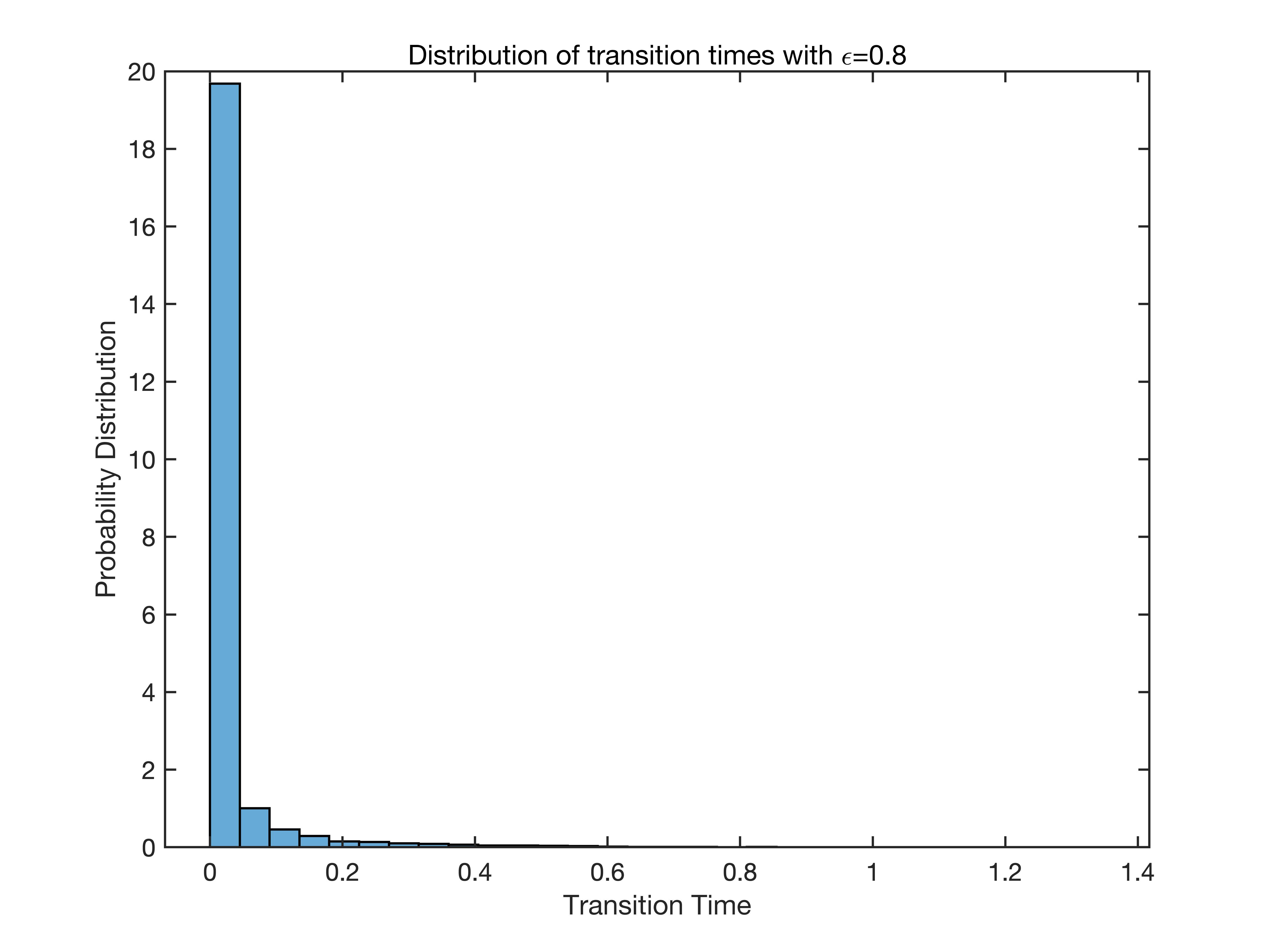}
    \end{minipage}
    \label{fig:es0.8}
}
\subfigure[$\epsilon=0.9$]{
    \begin{minipage}[b]{0.4\textwidth}
    \includegraphics[width=1\textwidth]{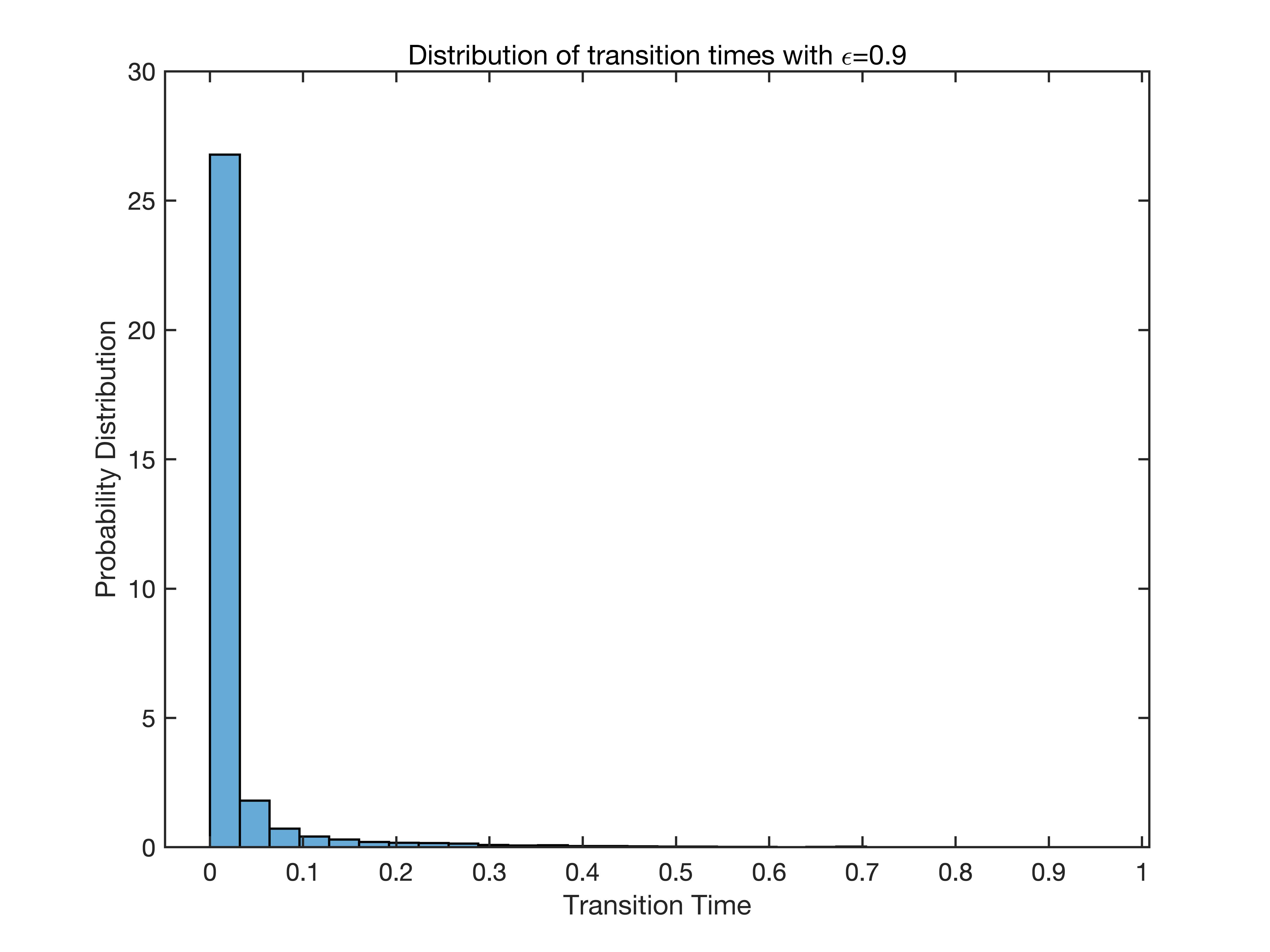}
    \end{minipage}
    \label{fig:es0.9}
}
\caption{The distribution of transition times with various noise coefficient. (a) $\epsilon=0.1$; (b) $\epsilon=0.3$;
(c) $\epsilon=0.5$;
(d) $\epsilon=0.7$; (e) $\epsilon=0.8$; (f) $\epsilon=0.9$.}
\label{fig:es}
\end{figure}

Utilizing Matlab, we perform simulations of Eq.(\ref{eq_stochastic}) for $N=10000$. The resultant distribution of the most probable transition times is depicted in Figure \ref{fig:es}. Here, the mean transition time signifies the arithmetic mean of all transition times observed. The corresponding mean transition times are as follows: (a) $0.1122$; (b) $0.0531$; (c) $0.0344$; (d) $0.0293$; (e) $0.0257$; (f) $0.0242$. It is evident that as the noise intensity increases, the peak of the sample point distribution gradually becomes more pronounced. Furthermore, regardless of the noise intensity, the majority of transition times are concentrated before $0.2$, with only a negligible fraction extending beyond $0.2$. This indicates that the system undergoes transitions between metastable states within an extremely short period of time. To maintain generality, we opt for $T=0.0344$ as the transition time, proceeding to solve the subsequent boundary value problem.

\subsection{The most probable transition pathway}
\indent

\begin{figure}[H]
\centering
\subfigure[$\epsilon=0.3$]{
    \begin{minipage}[b]{0.4\textwidth}
    \includegraphics[width=1\textwidth]{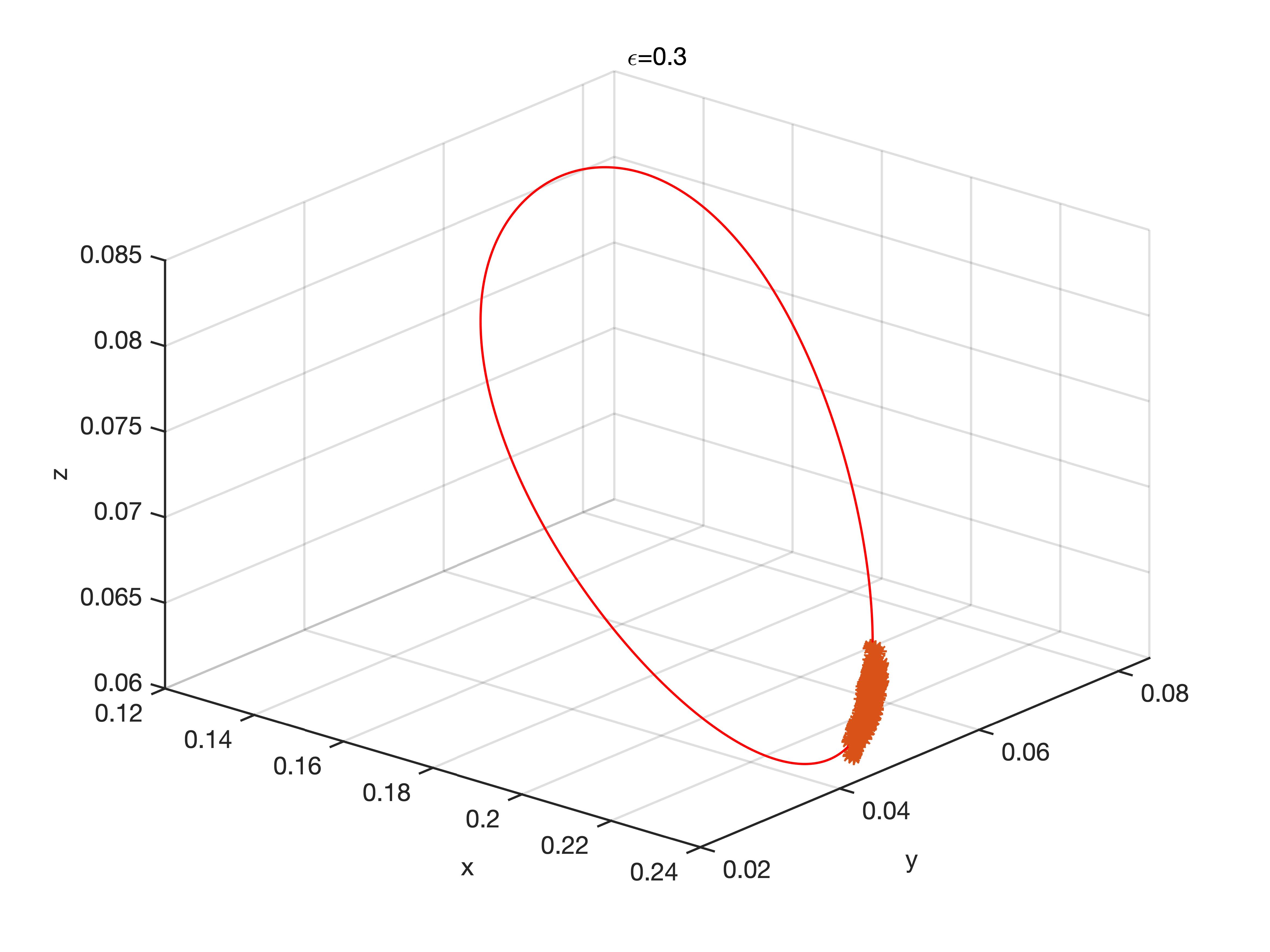} 
    \end{minipage}
    \label{fig:end0.3}
}
\subfigure[$\epsilon=0.5$]{
    \begin{minipage}[b]{0.4\textwidth}
    \includegraphics[width=1\textwidth]{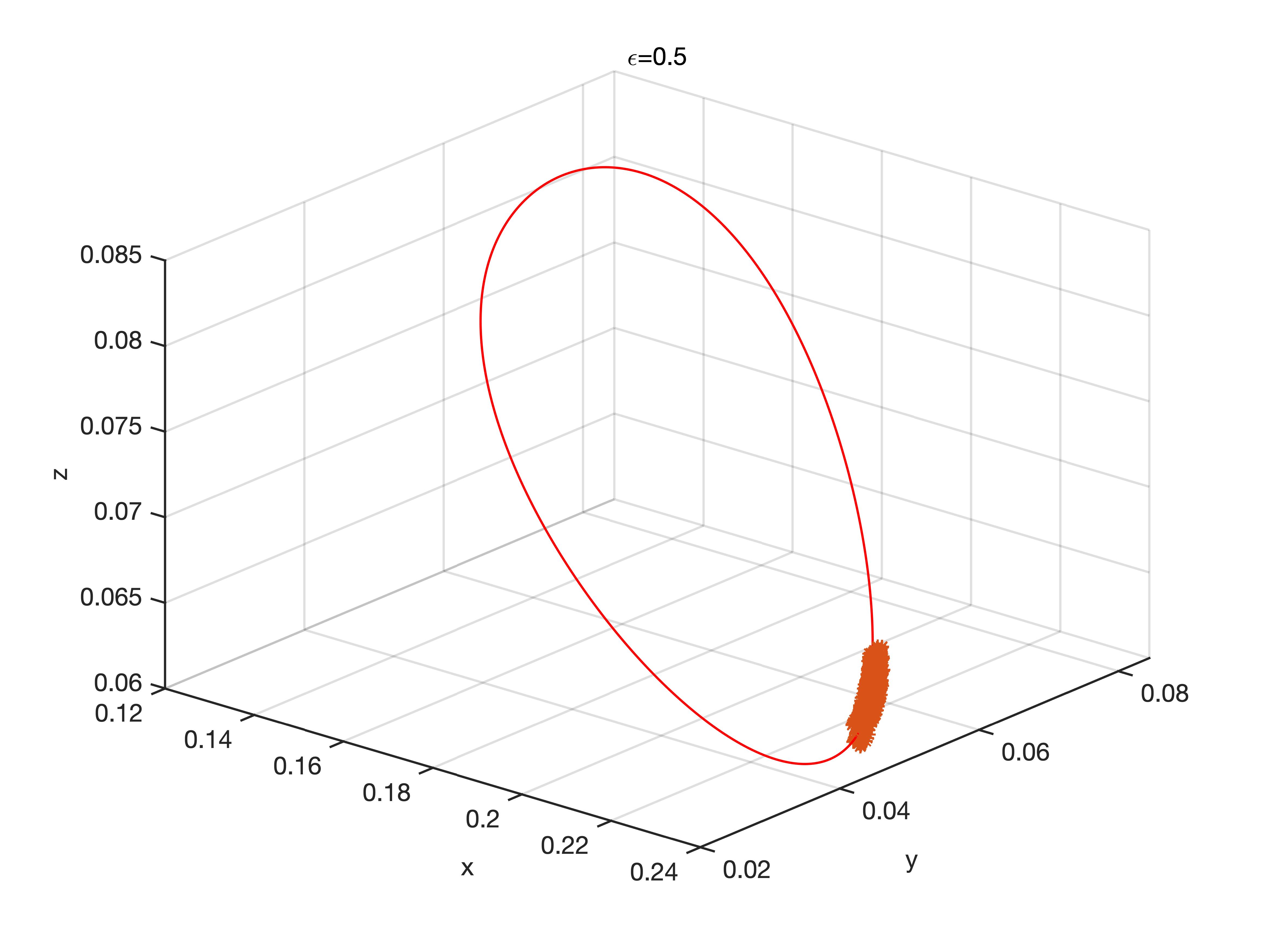}
    \end{minipage}
    \label{fig:end0.5}
}\\
 
\subfigure[$\epsilon=0.7$]{
    \begin{minipage}[b]{0.4\textwidth}
    \includegraphics[width=1\textwidth]{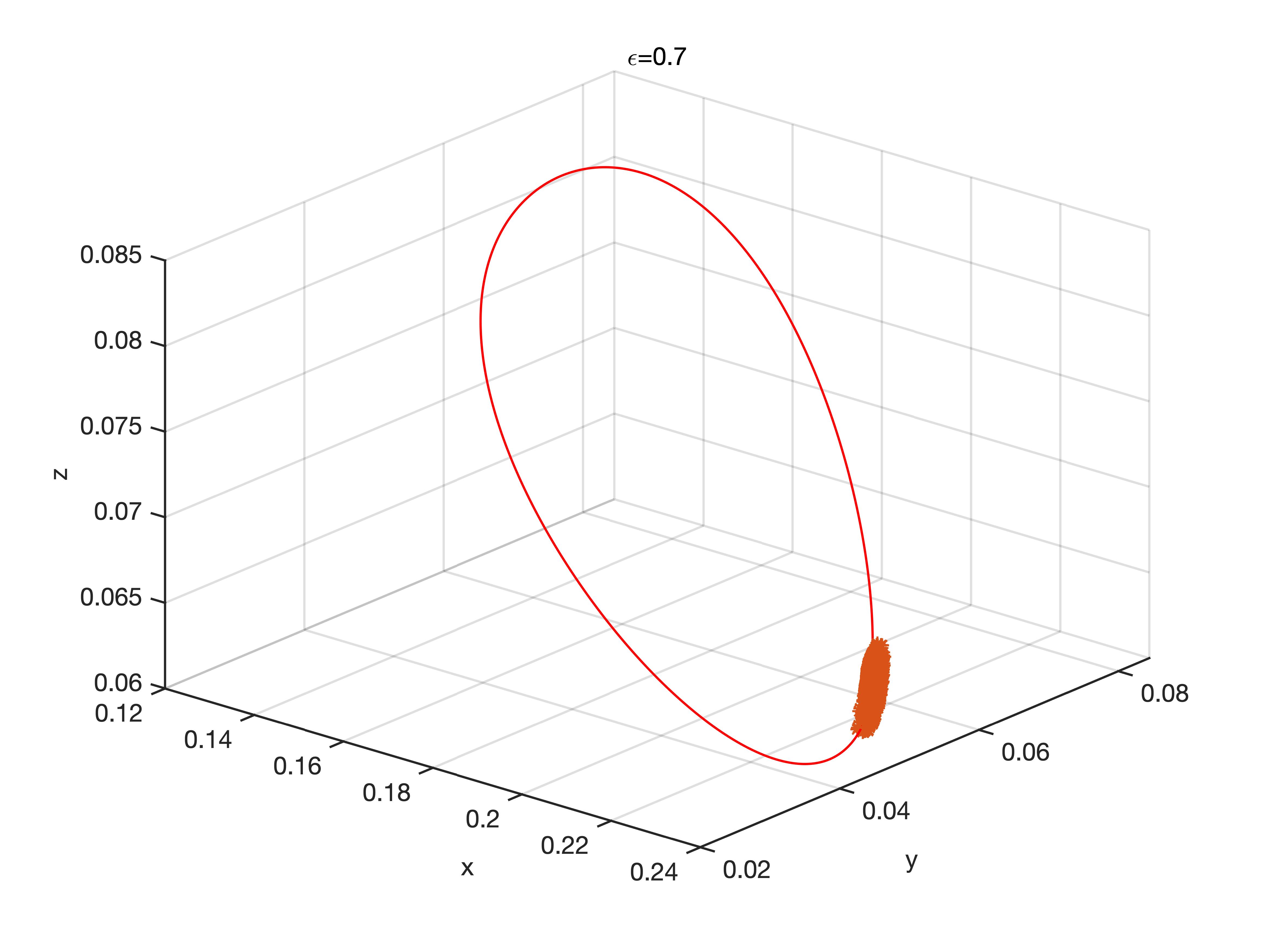}
    \end{minipage}
    \label{fig:end0.7}
}
\subfigure[$\epsilon=0.9$]{
    \begin{minipage}[b]{0.4\textwidth}
    \includegraphics[width=1\textwidth]{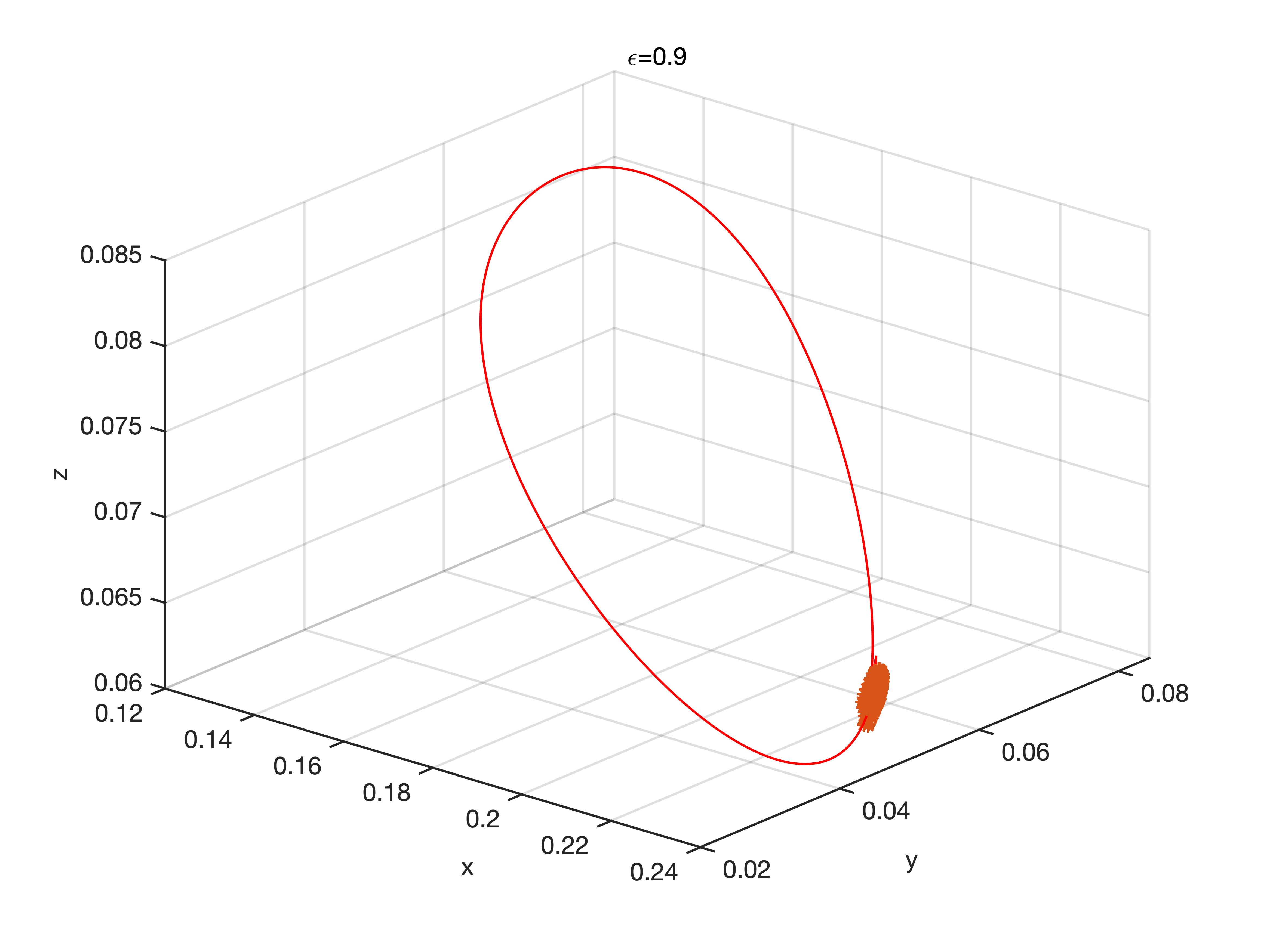}
    \end{minipage}
    \label{fig:end0.9}
}
\caption{The end point distribution of the simulated pathway with different noise coefficients. (a) $\epsilon=0.3$; (b) $\epsilon=0.5$;
(c) $\epsilon=0.7$; (d) $\epsilon=0.9$.}
\label{fig:end}
\end{figure}

Figure \ref{fig:end} depicts the distribution of sample points generated on the limit cycle. It is evident that the noise intensity affects the dispersion of the sample endpoints on the limit cycle. This indicates that when the noise intensity is higher, the most probable transition pathway of the system become more distinct. When comparing Figure \ref{fig:end0.3} with Figure \ref{fig:end0.9}, it is strikingly evident that as the noise intensity increases, the distribution of endpoints on the limit cycle becomes increasingly concentrated.

\begin{figure}[H] 
\centering
\subfigure[]{
    \begin{minipage}[b]{0.4\textwidth}
    \includegraphics[width=1\textwidth]{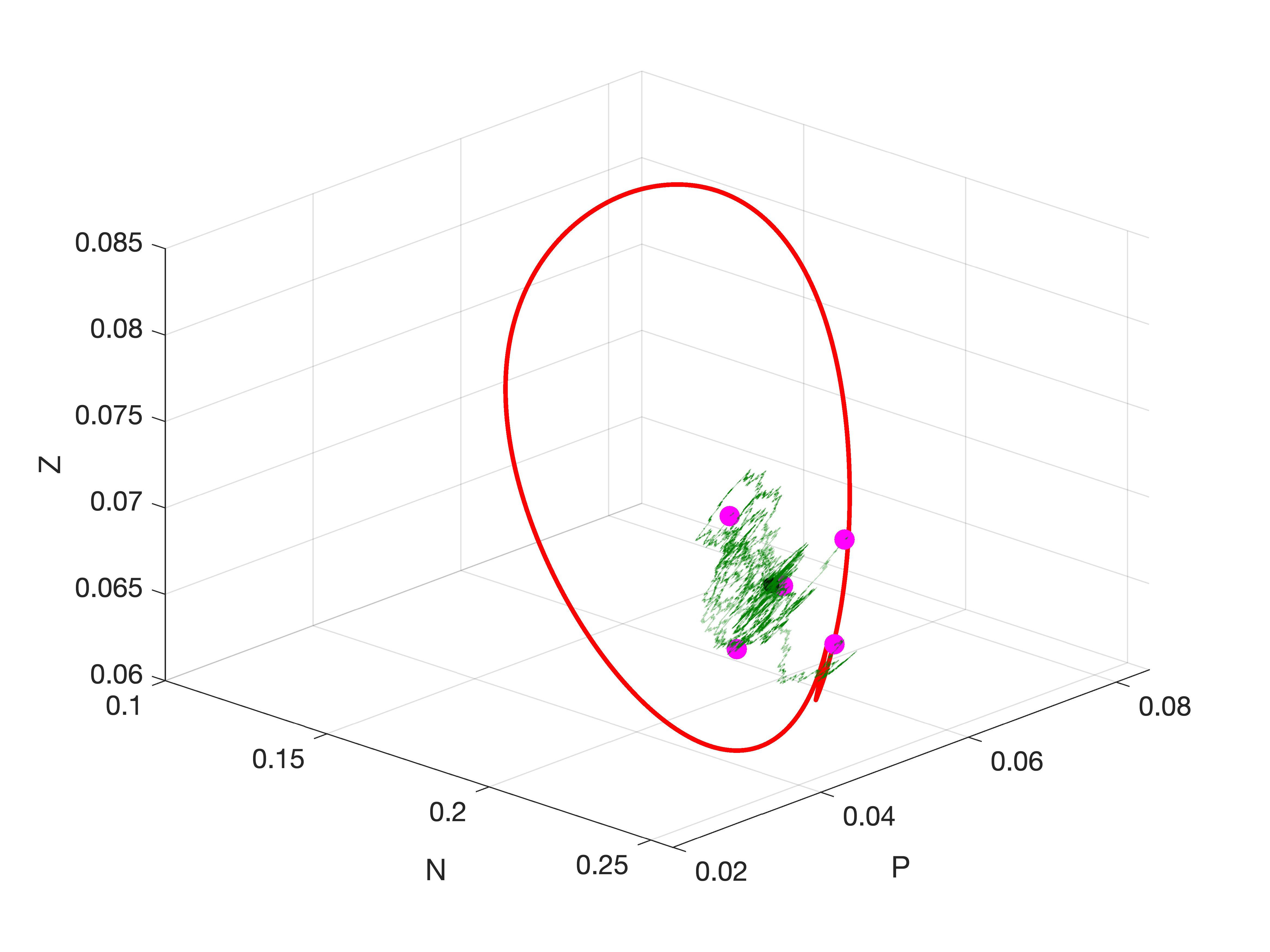} 
    \end{minipage}
    \label{fig:monte5}
}
\subfigure[]{
    \begin{minipage}[b]{0.4\textwidth}
    \includegraphics[width=1\textwidth]{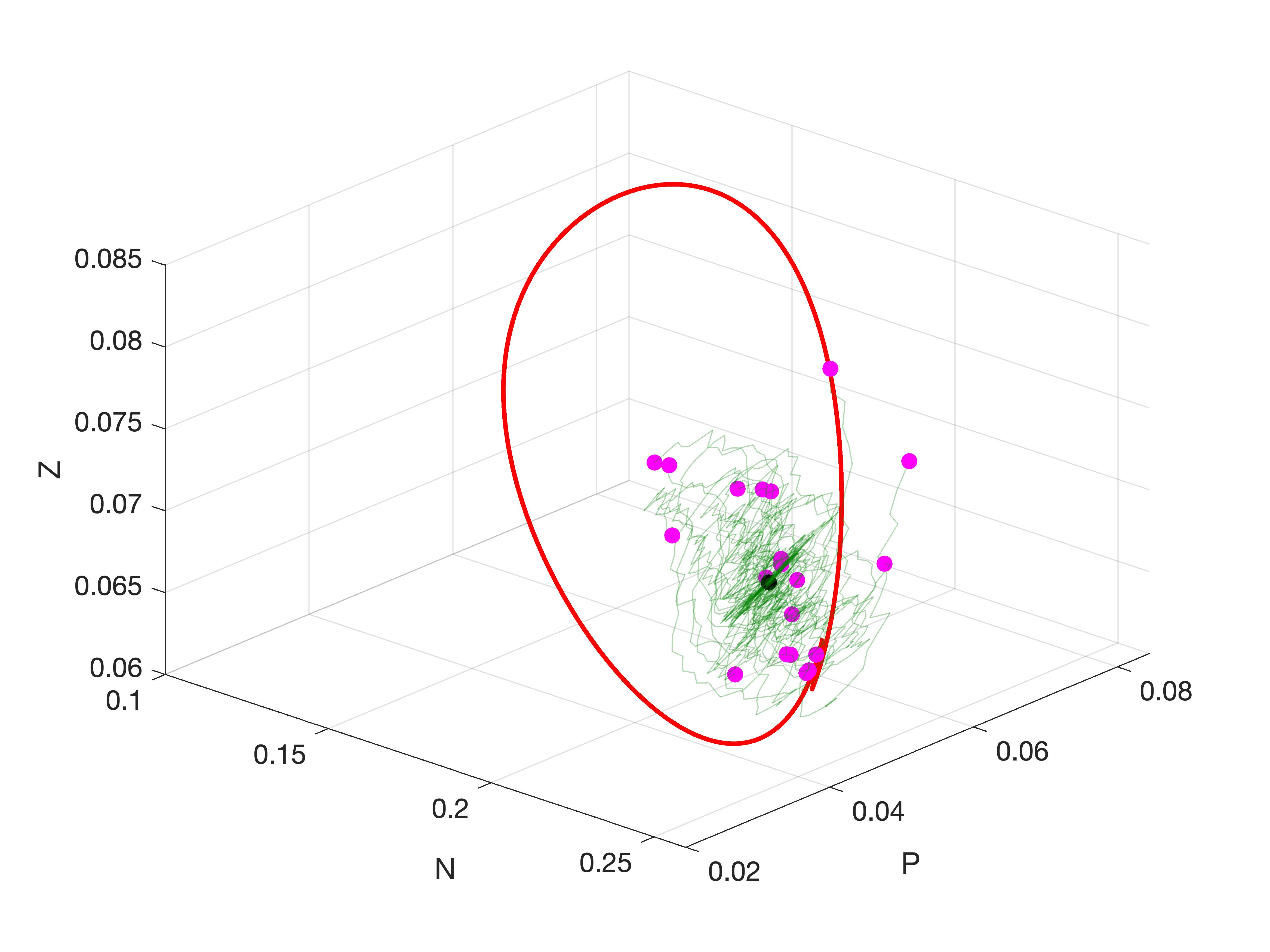}
    \end{minipage}
    \label{fig:monte20}
}\\
\subfigure[]{
    \begin{minipage}[b]{0.4\textwidth}
    \includegraphics[width=1\textwidth]{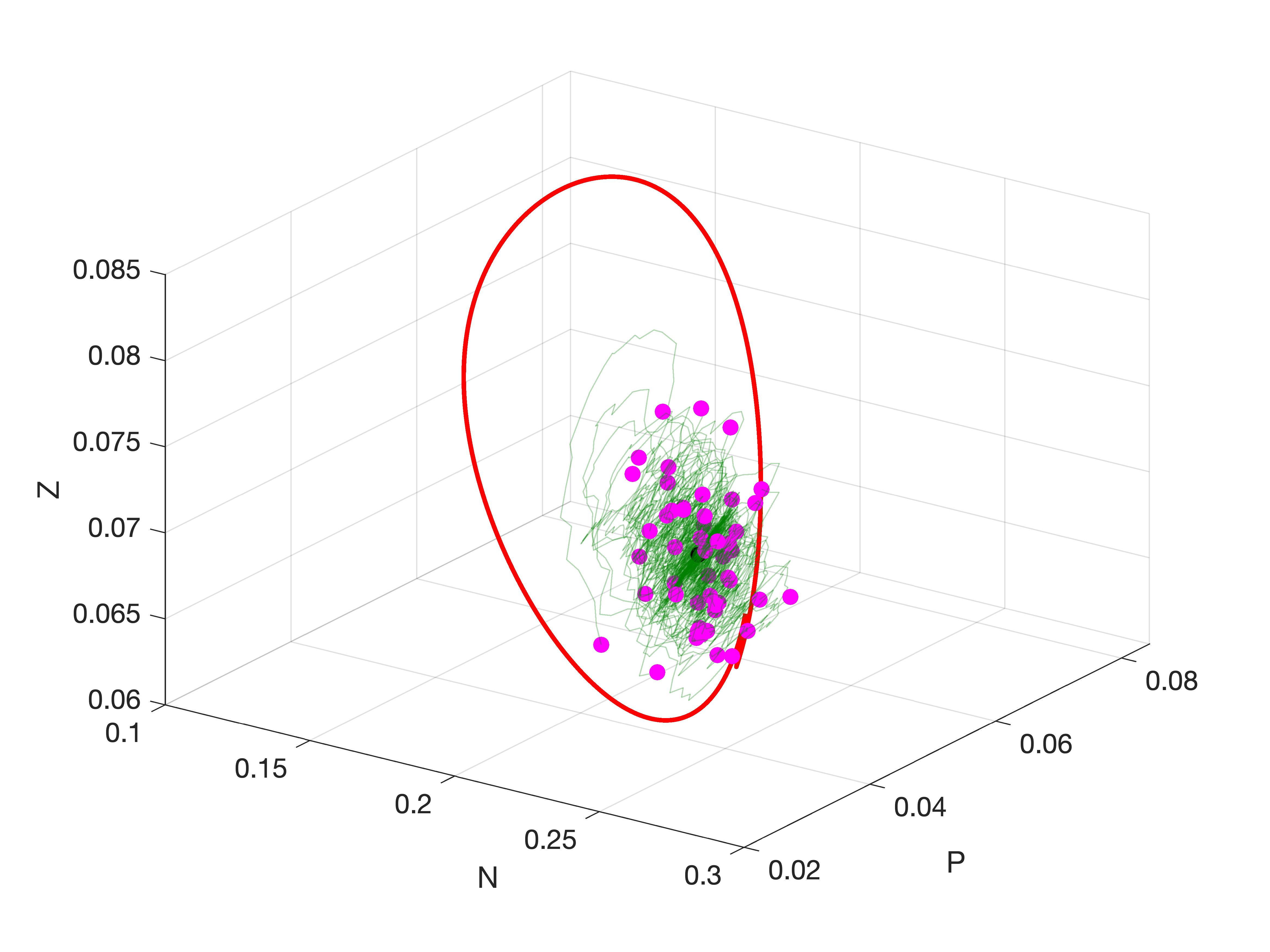}
    \end{minipage}
    \label{fig:monte50}
}
\subfigure[]{
    \begin{minipage}[b]{0.4\textwidth}
    \includegraphics[width=1\textwidth]{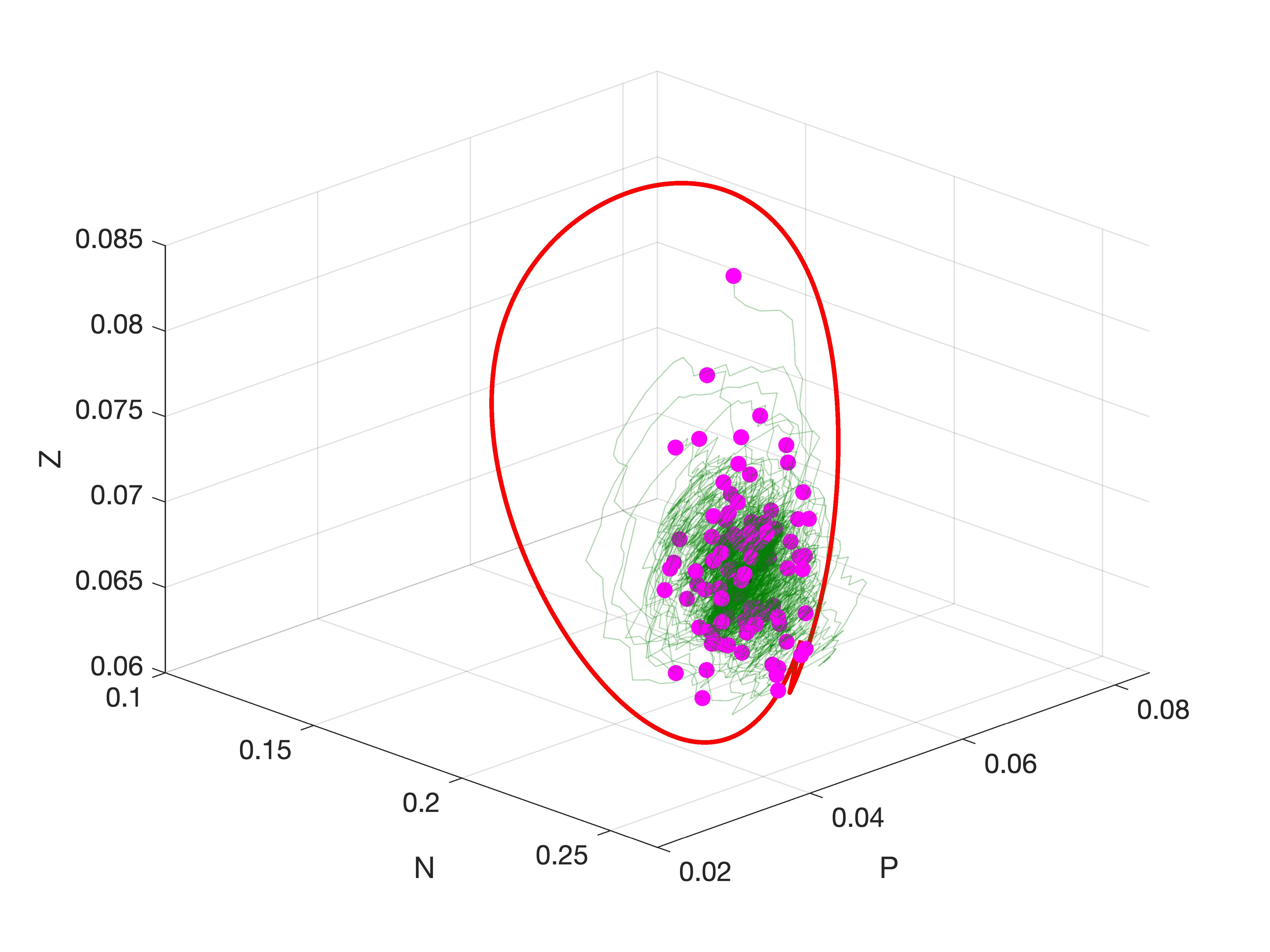}
    \end{minipage}
    \label{fig:monte100}
}
\caption{The simulated paths based on Eq.(\ref{eq_deter}) are shown in the figure. (a) $5$ paths; (b) $20$ paths;
(c) $50$ paths;
(d) $100$ paths.}
\label{fig:monte}
\end{figure}

Figure \ref{fig:monte} presents the simulated paths of the NPZ system using the Monte Carlo method. The red line represents the limit cycle of the system, the green lines depict the simulated paths, and the purple dots mark the endpoints of these paths. We conduct simulations for $5$, $20$, $50$, and $100$ paths. It can be observed from Figure \ref{fig:monte} that as the number of simulated paths increases, so does the number of paths that successfully reach the limit cycle. Moreover, most of the paths that do reach the limit cycle tend to converge towards the lower right section of the cycle, which corresponds to the endpoint distribution observed in Figure \ref{fig:end}. This provides a basis for investigating the most probable transition pathways.

\begin{figure}[H]
\centering
\subfigure[$\epsilon=0.3$]{
    \begin{minipage}[b]{0.4\textwidth}
    \includegraphics[width=1\textwidth]{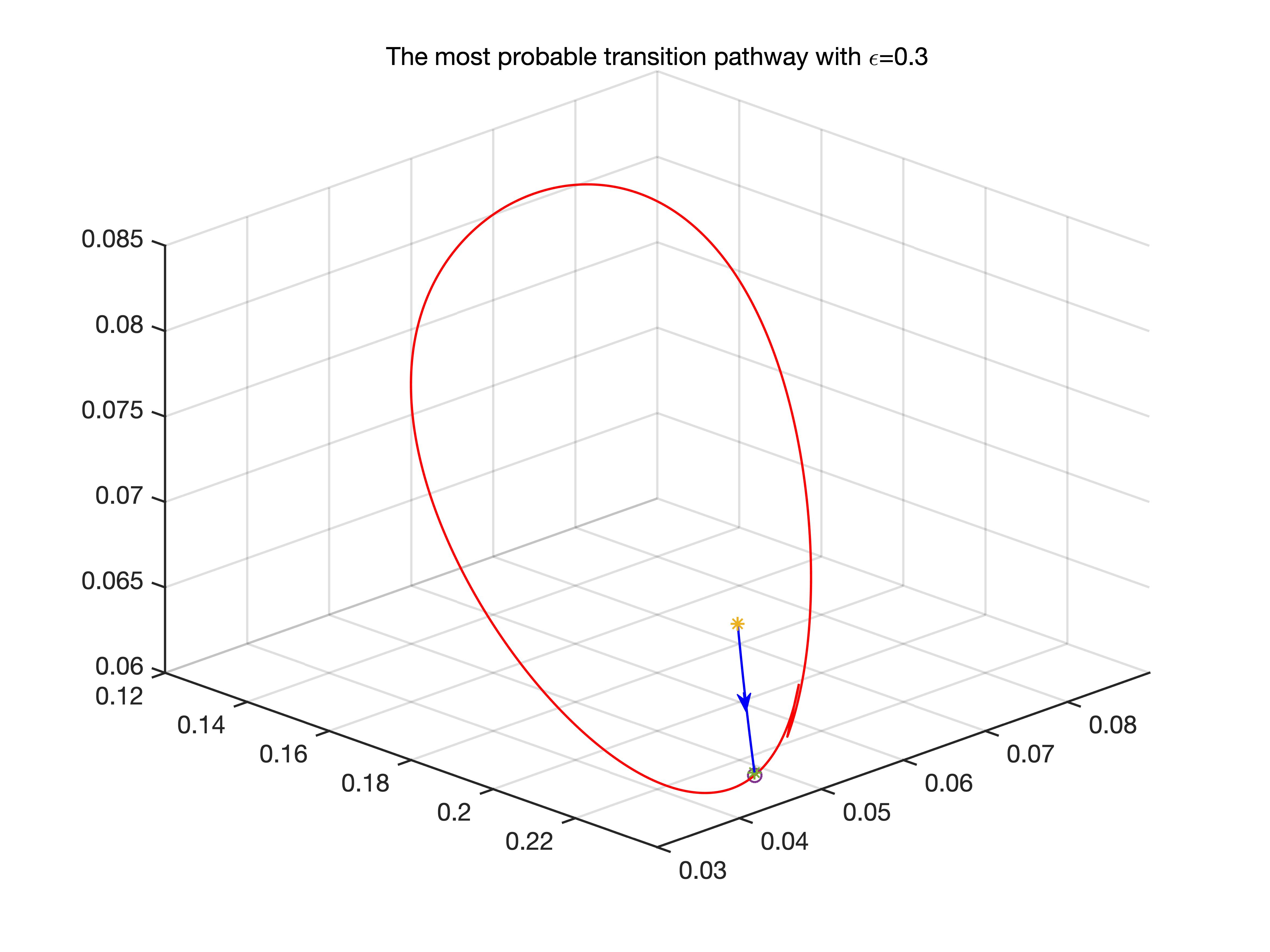} 
    \end{minipage}
    \label{fig:path0.3}
}
\subfigure[$\epsilon=0.5$]{
    \begin{minipage}[b]{0.4\textwidth}
    \includegraphics[width=1\textwidth]{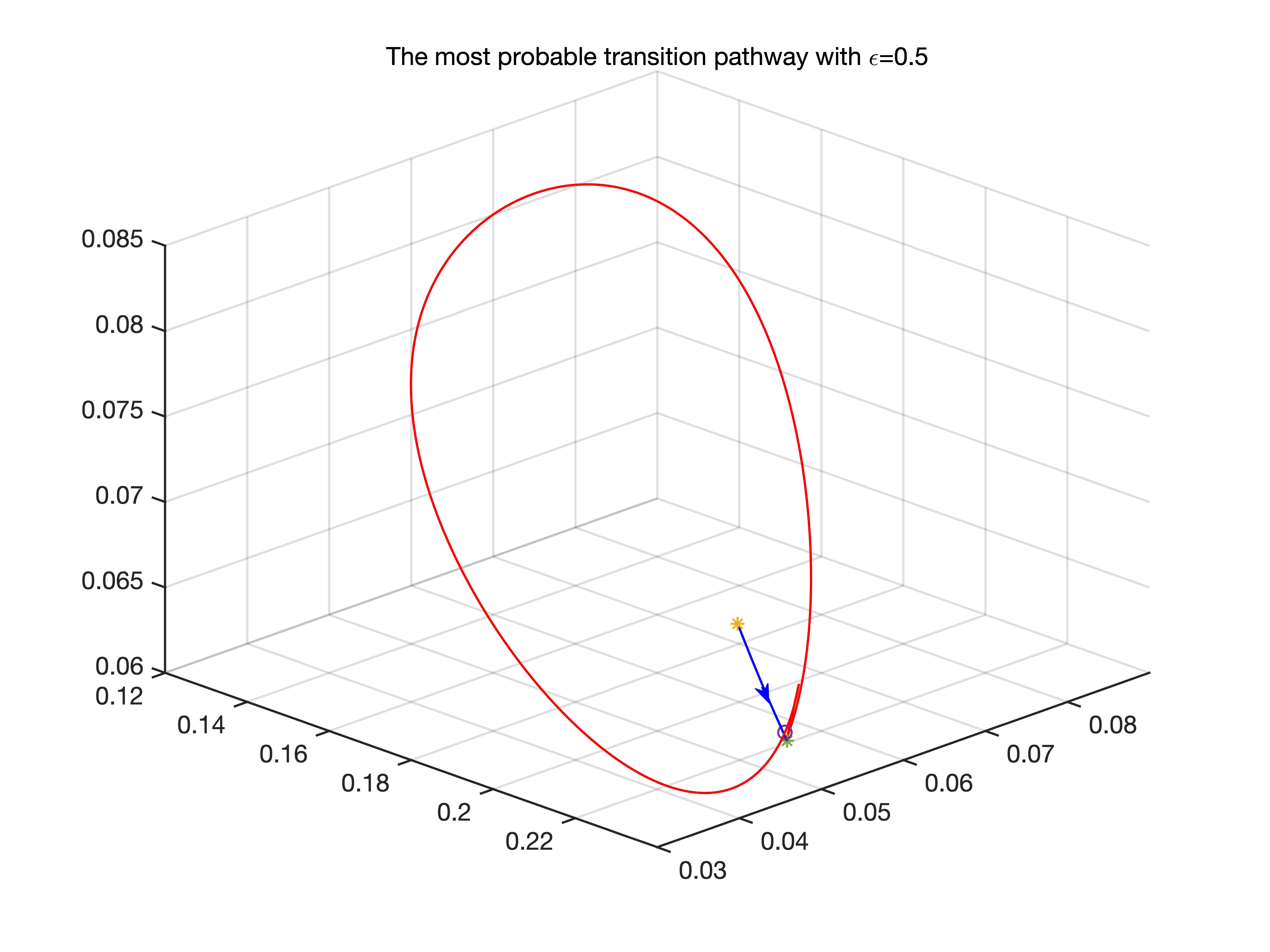}
    \end{minipage}
    \label{fig:path0.5}
}\\
 
\subfigure[$\epsilon=0.7$]{
    \begin{minipage}[b]{0.4\textwidth}
    \includegraphics[width=1\textwidth]{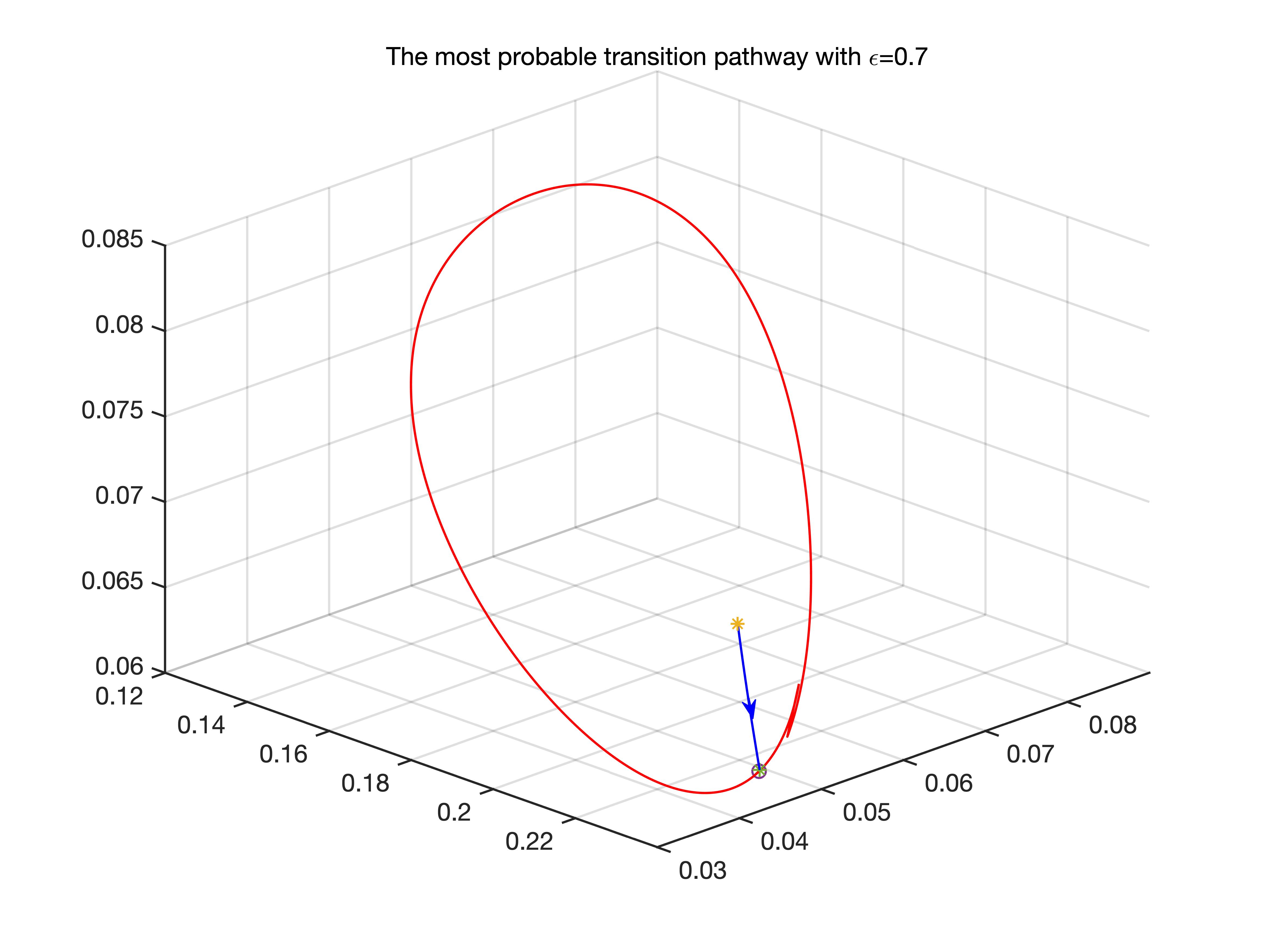}
    \end{minipage}
    \label{fig:path0.7}
}
\subfigure[$\epsilon=0.9$]{
    \begin{minipage}[b]{0.4\textwidth}
    \includegraphics[width=1\textwidth]{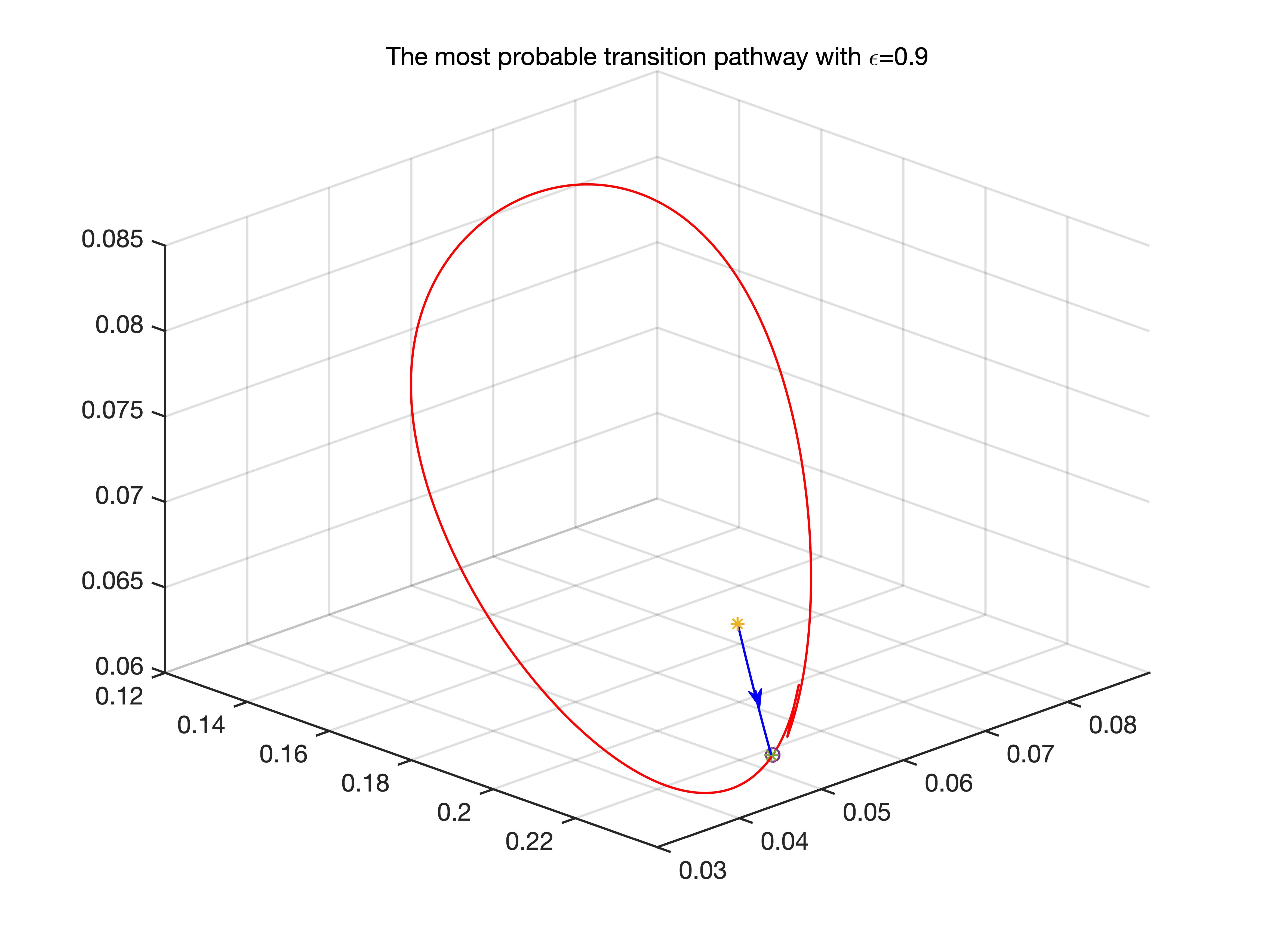}
    \end{minipage}
    \label{fig:path0.9}
}\\

\subfigure[$\epsilon=0.1$]{
    \begin{minipage}[b]{0.4\textwidth}
    \includegraphics[width=1\textwidth]{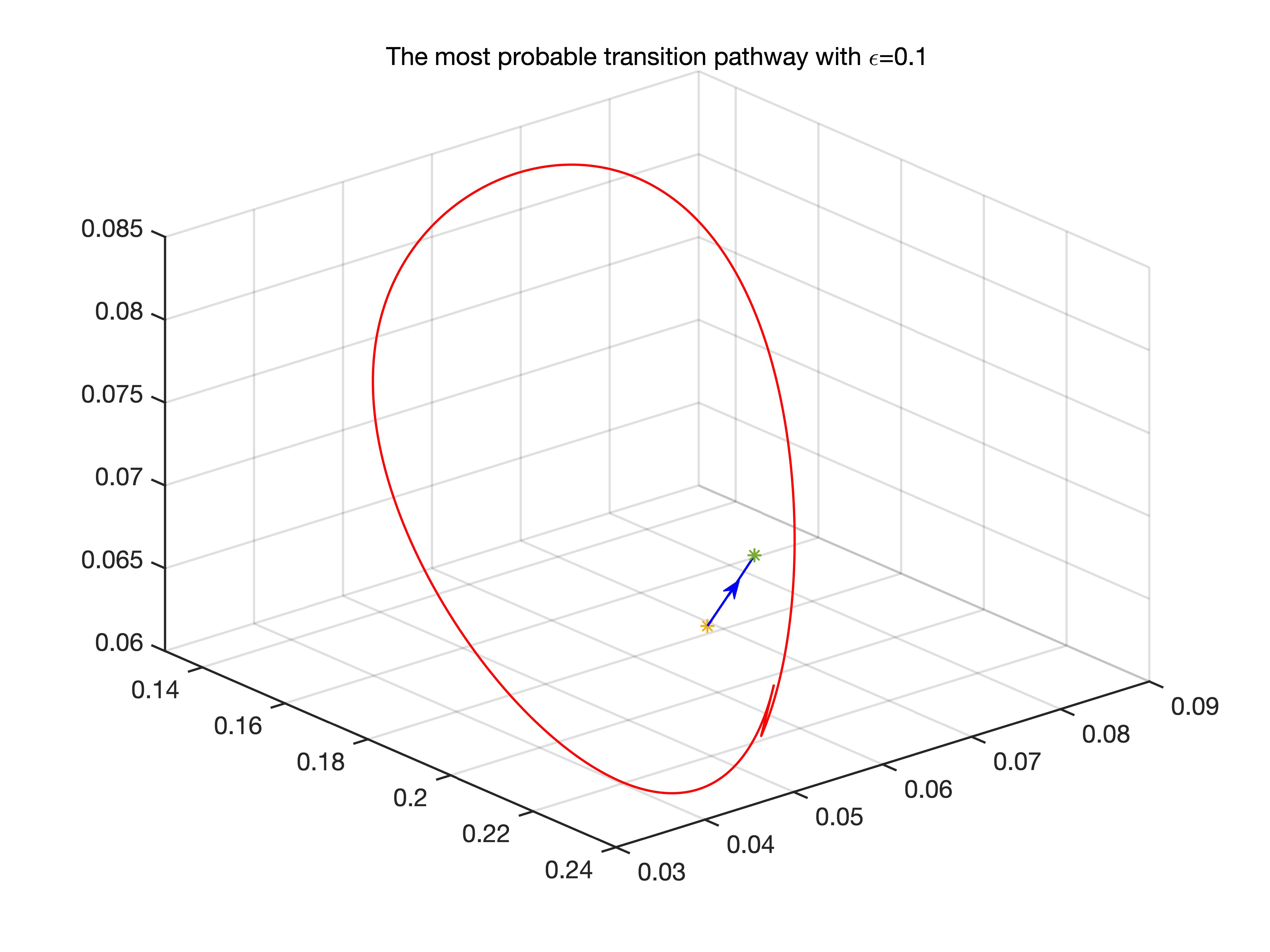}
    \end{minipage}
    \label{fig:not0.1}
}
\subfigure[$\epsilon=0.5$]{
    \begin{minipage}[b]{0.4\textwidth}
    \includegraphics[width=1\textwidth]{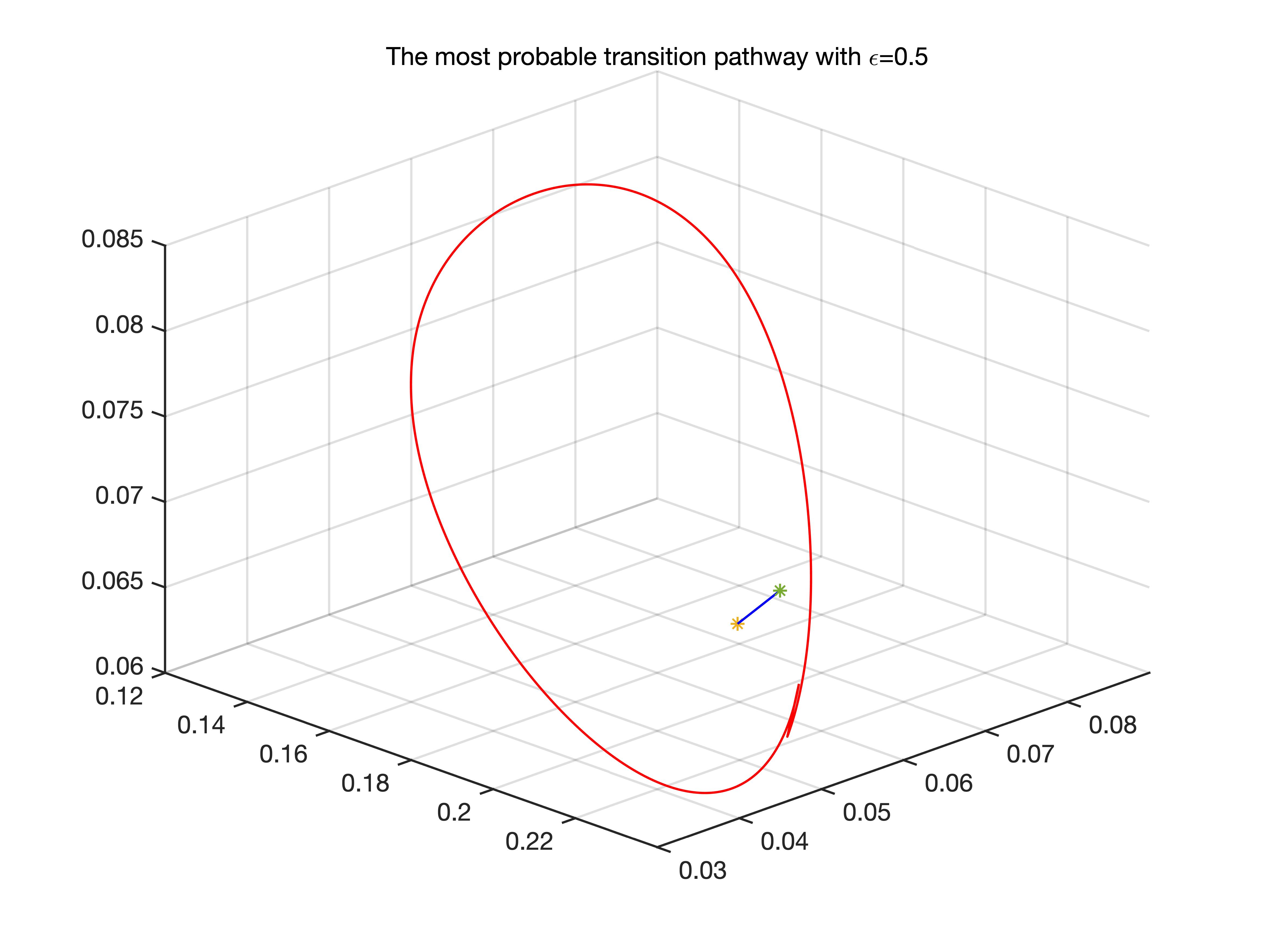}
    \end{minipage}
    \label{fig:not0.5}
}\\

\caption{The most probable transition pathways with different noise coefficients. (a) $\epsilon=0.3$; (b) $\epsilon=0.5$;
(c) $\epsilon=0.7$;
(d) $\epsilon=0.9$. The unreachable pathway: (e) $\epsilon=0.1$;
(f) $\epsilon=0.5$.}
\label{fig:path}
\end{figure}

We select the mean transition time at $\epsilon=0.5$ as the transition time $T$ of the system through the distribution of transition time under different noise intensities. As shown in Figure \ref{fig:es}, noise intensity affects the transition time between metastable states of the NPZ system. Figure \ref{fig:path} illustrates the most probable transition pathways of the system from the equilibrium point to the limit cycle under various noise coefficients. The red ellipse represents the limit cycle, and the blue arrows indicate the pathway of particle transitions. The first four figures depict the pathways that the system can transition to the limit cycle under the influence of noise intensity, whereas the last two figures show scenarios where the system cannot reach the limit cycle. As the noise coefficient increases, the transition pathways of the particles become more complex and unpredictable. Since the distribution of end points on the limit cycle is concentrated in the lower right region of the limit cycle, the system will transition to a specific area on the limit cycle when starting from an equilibrium point. By comprehending these dynamic relationships, we can understand that appropriate environmental fluctuations can promote the survival and reproduction of populations, while also influencing the evolution of other species. This phenomenon enables us to better study and predict the impact of environmental changes on ecosystems, which is of great significance for the management and protection of ecosystems.

However, not all pathways can precisely reach the limit cycle. The most probable transition pathways are typically derived from the initial velocities estimated by neural networks. Even if we increase the sample size and restrict the training error, there will still be pathways that fail to meet the necessary conditions, with their endpoints not on the limit cycle. We refer to these pathways as ``unreachable pathways". Figure \ref{fig:not0.1} and Figure \ref{fig:not0.5} illustrate the transition pathway that fail to reach the limit cycle under the same conditions.

\begin{figure}[H]
\centering
\subfigure[]{
    \begin{minipage}[b]{0.4\textwidth}
    \includegraphics[width=1\textwidth]{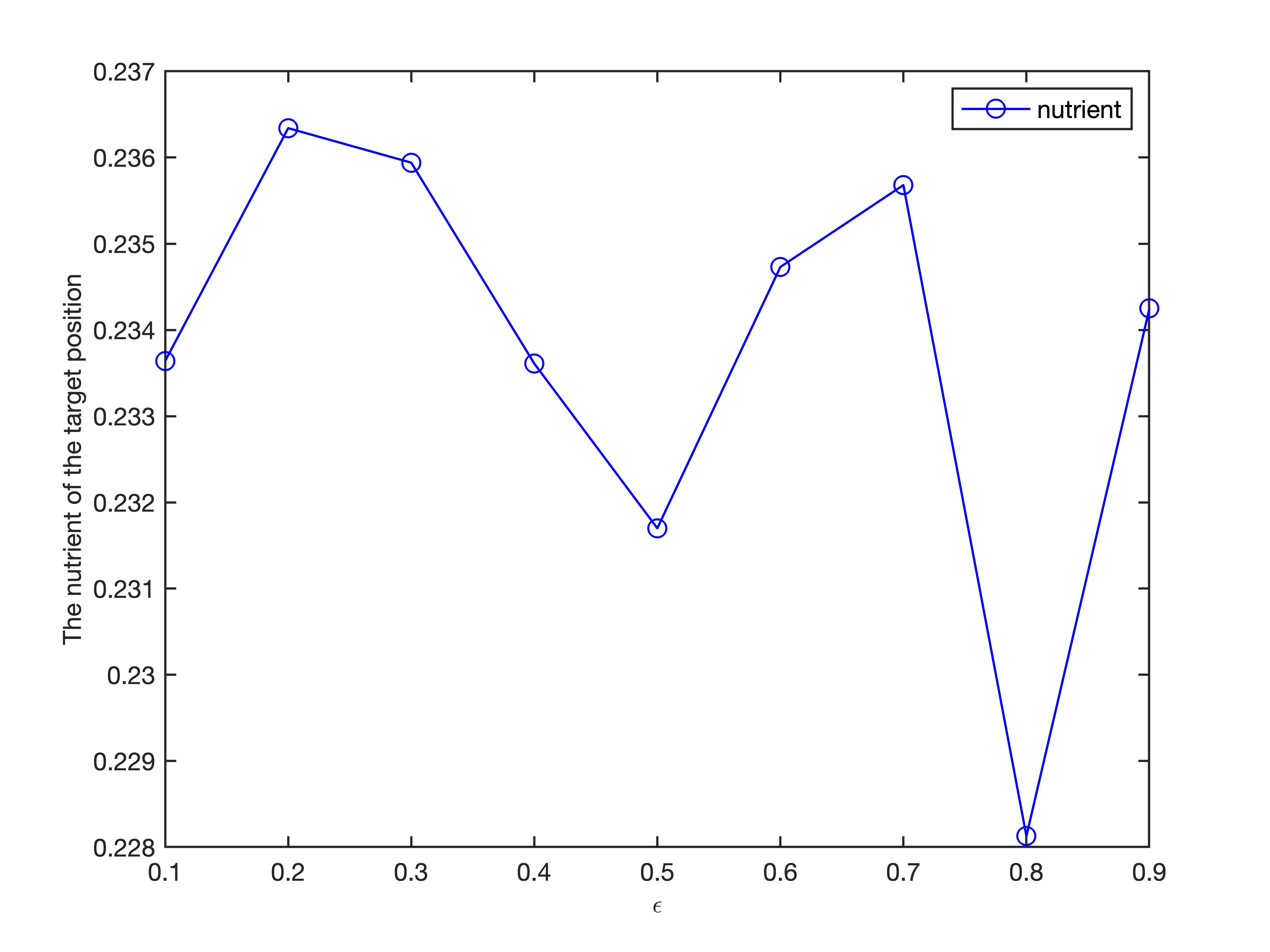} 
    \end{minipage}
    \label{fig:tp1}
}\\
\subfigure[]{
    \begin{minipage}[b]{0.4\textwidth}
    \includegraphics[width=1\textwidth]{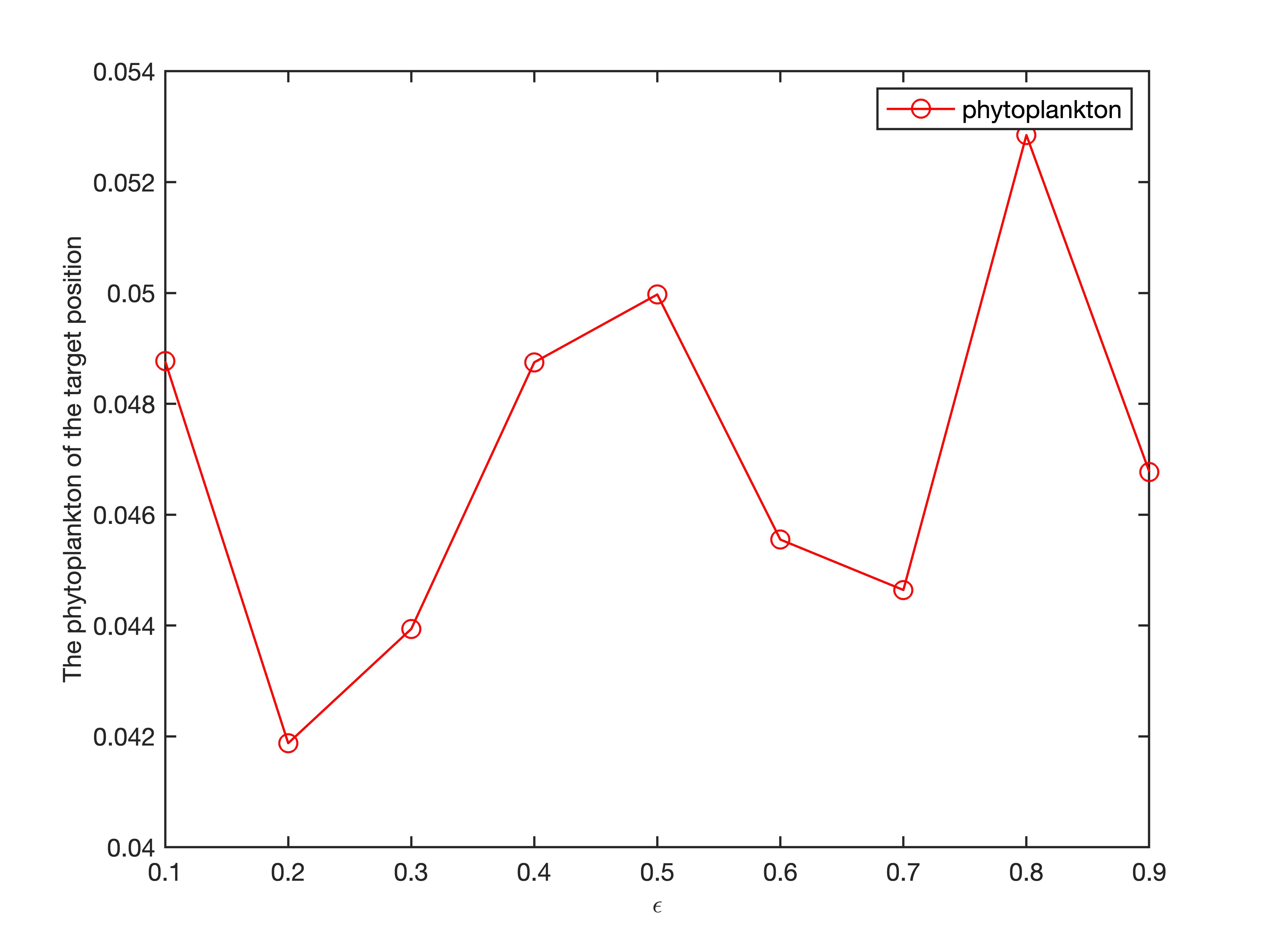}
    \end{minipage}
    \label{fig:tp2}
}\\
\subfigure[]{
    \begin{minipage}[b]{0.4\textwidth}
    \includegraphics[width=1\textwidth]{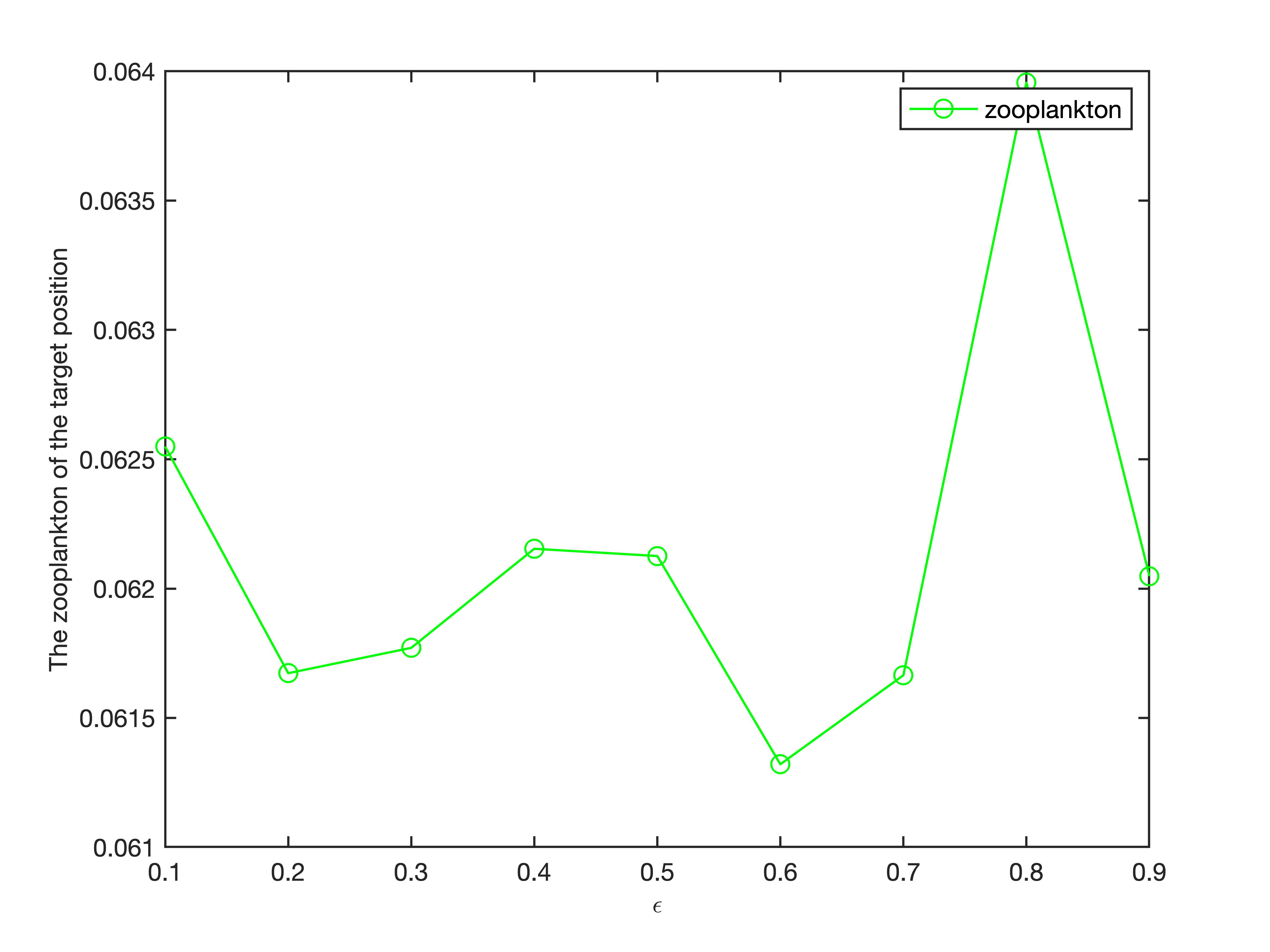}
    \end{minipage}
    \label{fig:tp3}
}
\caption{The target positions of the most probable transition pathway of nutrient, phytoplankton and zooplankton with different noise coefficients. (a) nutrient; (b) phytoplankton; (c) zooplankton.}
\label{fig:tp}
\end{figure}

Figure \ref{fig:tp} depicts the most probable target positions of the transition pathways for nutrients, phytoplankton and zooplankton under varying noise coefficients. As observed from Figure \ref{fig:tp}, the target positions of these three biological components fluctuate with changes in noise intensity, exhibiting distinct amplitudes and trends. The number of nutrients initially increases and then decreases, with increasing noise intensity causing the nutrient levels to oscillate up and down. Phytoplankton, in contrast, first experiences a decline, followed by a gradual increase, and ultimately oscillates. Zooplankton follows a similar trend to phytoplankton, but its amplitude of change is smaller under low noise intensity. However, when the noise intensity exceeds $0.7$, the fluctuation of zooplankton becomes significantly greater. 

Under high noise levels, the populations of phytoplankton and zooplankton tend to increase, while the population of nutrients decreases. This indicates that phytoplankton are actively utilizing nutrients for growth and reproduction. At the same time, the population of zooplankton also increases because they feed on phytoplankton. However, as the number of zooplankton increases, the population of phytoplankton may decline due to over-predation, which in turn affects the population of zooplankton. This negative feedback mechanism helps maintain the stability of the ecosystem in the face of environmental changes. Similarly, under low noise levels, the populations of phytoplankton and zooplankton decrease, while the availability of nutrients increases. This dynamic relationship reflects ecosystems' inherent capacity for self-regulation through internal interactions. Our findings fundamentally reshape the understanding of ecosystem stability and dynamics, elucidating how biodiversity and ecological systems undergo abrupt changes in response to external disturbances. These insights significantly enhance our ability to investigate and predict ecosystem responses to environmental changes.

\subsection{The most probable transition probability}
\indent

\begin{figure}[H]
    \centering
    \includegraphics[width=0.5\linewidth]{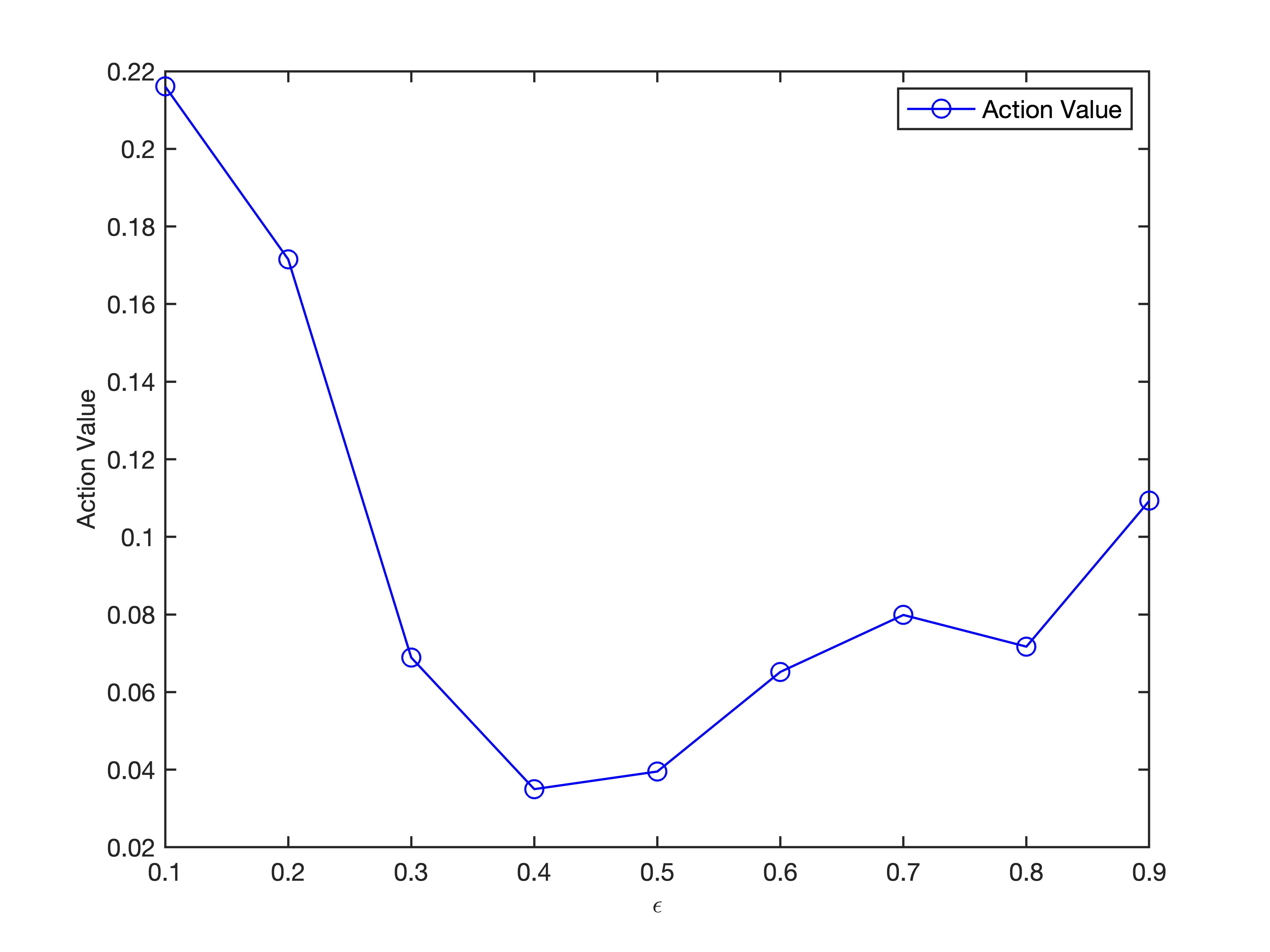}
    \caption{The action value of the different noise intensities.}
    \label{fig:action}
\end{figure}

Based on the Onsager-Machlup action functional
$$P(\left \| X-\omega \right \|_{t} \le \delta)\propto C(\delta,t)exp \left\{ -S(\omega,\dot{\omega})\right\},$$
we can find that the minimum action functional value corresponds to the maximum probability value.

Figure \ref{fig:action} illustrates the values of the action functional for varying noise intensities. Each noise intensity is associated with a minimum action functional value. The maximum probability value represents the most probable transition probability of the three-dimensional NPZ system. The figure shows that as the noise intensity fluctuates, the evolution of species becomes increasingly complex and unstable. This is because the actual environment is replete with a wide array of intricate and diverse perturbation factors, all of which exert varying degrees of influence on species. In particular, the action functional value initially decreases sharply with increasing noise intensity. After the noise intensity reaches $0.4$, the value begins to gradually increase and eventually exhibits minor fluctuations between $\epsilon=0.7$ and $\epsilon=0.9$. 

When the parameters $\delta$ and $T$ are fixed, it can be observed that $P$ depends solely on the action functional value, while the coefficient $C(T, \delta)$ can be regarded as a constant that does not vary with the final position on the limit cycle. Since our objective is to qualitatively explore the differences in transition probabilities to each segment, we can disregard the constant $C(T, \delta)$ to obtain an approximate probability of the most probable transition pathway for each segment. Therefore, we can deduce that as the noise intensity increases, the most probable transition probability first increases and then decreases, with slight variations occurring when the noise intensity exceeds $0.7$.

The rate of decline in the action value suddenly increases at $\epsilon=0.2$; at $\epsilon=0.8$, it surges sharply. These peculiarities in the intensity of noise are also observed in the aforementioned analysis. This indicates that an increase in the population of phytoplankton and zooplankton leads to a decrease in the action value, while the density of nutrients may reach its maximum. Conversely, at $\epsilon=0.8$, the density of nutrients increases while the populations of the other two species decrease, and the densities of phytoplankton and zooplankton reach their peaks. This further illustrates that, under the influence of complex external environmental disturbances, species can maintain the relative stability of the entire ecosystem through mutual promotion and mutual restraint.

\section{Discussion}
\indent

The phenomenon of transitions between metastable states induced by stochastic noise holds great significance in ecosystems. We focuses primarily on the nutrient-phytoplankton-zooplankton system in marine ecological environments. Firstly, we analyze the stability of the equilibrium points and limit cycle of the system using the Jacobian matrix and Lyapunov exponents. Subsequently, based on the Onsager-Machlup action functional and the neural shooting method, we examine the transitions between metastable states in the NPZ system under the influence of noise. Additionally, we employ the Monte Carlo method to simulate the stochastic paths of the system, thereby verifying the direction of the most probable transition pathways. Specifically:

$\bullet$ For the deterministic NPZ system, when the predation rate $d$ takes an appropriate value, the system has one stable equilibrium point and one stable limit cycle.

$\bullet$ Different noise intensities affect the most probable transition time of the system. We find that the most probable transition times are mostly concentrated within $0.2$. Moreover, the mean transition time decreases as the noise intensity increases.

$\bullet$ When employing the Monte Carlo method to simulate the stochastic paths of the system, it is observed that the generated sample points are predominantly clustered in the lower right region of the limit cycle. Interestingly, the concentration of these sample points intensifies with increasing noise intensity.

$\bullet$ Noise-induced perturbations can prompt the system to transition from a stable equilibrium point to a stable limit cycle. Notably, variations in noise intensity lead to changes in the target position of the most probable transition pathway. This target position, in turn, reveals the intricate interplay of mutual promotion and restraint among the three species.

$\bullet$ Based on the Onsager-Machlup action functional theory, the minimum action functional value corresponds to the most probable transition probability. The most probable transition probability of the NPZ system decreases rapidly with increasing noise intensity, then increases slowly, and finally exhibits slight fluctuations.

This study investigates the most probable transition phenomenon and analyzes how the system transition between two metastable states under the noise perturbations. This transition phenomenon enhances our understanding of the dynamical behavior of plankton systems. Our research contributes to elucidating the underlying mechanisms governing population fluctuations in ecosystems. It provides profound insights into the mutual constraints among species in marine ecosystems, while enabling predictive assessments of environmental impacts on ecological systems. This work facilitates early prevention of detrimental ecological regime shifts and establishes a theoretical foundation with practical implications for ecosystem management.

\section*{Acknowledgments}

We would like to thank Miaolei Zheng for helpful discussions. This work is supported by the National Natural Science Foundation of China (NSFC) (Grant No.12101473).

\section*{Data availability statement}
No data was used for the research described in the article.

\section*{Declaration of competing interest}
The authors declare that they have no known competing financial interests or personal relationships that could have appeared to influence the work reported in this paper.

\bibliographystyle{unsrt}

\bibliography{refs}

\end{document}